\def\ksl{\not{\hbox{\kern-2.3pt $k$}}}
\documentclass[preprint,showpacs,preprintnumbers,amsmath,amssymb]{revtex4}
\usepackage{graphicx}% Include figure files
\usepackage{dcolumn}% Align table columns on decimal point
\usepackage{bm}% bold math

\newcommand{\comment}[1]{}
\newcommand{\bl}{{[\![}}
\newcommand{\br}{{]\!]}}
\newcommand{\be}{\begin{equation}}
\newcommand{\ee}{\end{equation}}
\def\nnn{\nonumber}
\def\fc#1#2{{\frac{#1}{#2}}}
\def\lf{\left}\def\ri{\right}
\def\ts{\textstyle}
\def\h{\fc{1}{2}}
\def\req#1{(\ref{#1})}
\def\ap{\alpha'}
\def\im{{\rm Im}}
\def\ie{{\it i.e.\ }}
\def\p{\partial}
\def\vev#1{\langle #1 \rangle}
\def\Kc{{\cal K}}
\def\Tr{{\rm Tr}}
\def\F#1#2{_#1F_#2}
\def\FF#1#2{F\lf[{#1\atop #2}\ri]}
\def\ds{\displaystyle}
\def\cf{{\it c.f.\ }}
\def\ie{{\it i.e.\ }}

\def\ra{\rightarrow}
\def\lra{\longrightarrow}

\begin{document}

\preprint{\tt hep-th/0609175}\preprint{LMU--ASC 50/06}

\title{Multi--Gluon Scattering in Open Superstring Theory}

\author{Stephan Stieberger}
\affiliation{
Arnold--Sommerfeld--Center for Theoretical Physics,\\
Department f\"ur Physik,
Ludwig--Maximilians--Universit\"at M\"unchen,\\
Theresienstra\ss e 37, 80333 M\"unchen, Germany}
% \textbackslash\textbackslash

\author{Tomasz R. Taylor}
\affiliation{
Department of Physics, Northeastern University\\
Boston, MA 02115, United States of America
}%

\date{September, 2006}% It is always \today, today,
\begin{abstract}
We discuss the amplitudes describing $N$-gluon scattering 
in type I superstring theory, on a disk world-sheet. After reviewing the 
general structure
of amplitudes and the complications created by the presence of a large
number of vertices at the boundary, we focus on the most promising
case of maximally helicity violating (MHV) configurations because in
this case, the zero Regge slope limit ($\alpha'\to 0$) is particularly
simple. 
We obtain the full-fledged MHV disk amplitudes for $N=4,5$ and
$N=6$ gluons,  expressed in terms of one, two and six functions of kinematic
invariants, respectively. 
These functions represent certain boundary integrals - generalized Euler integrals - 
which for $N\geq 6$ 
correspond to multiple
hypergeometric series (generalized Kamp\'e de F\'eriet functions).
Their $\alpha'$-expansions lead to Euler-Zagier sums.
For arbitrary $N$, we  show that the
leading string corrections to the Yang-Mills amplitude, of order
${\cal O}(\alpha'^2)$, originate 
from the well-known $\alpha'^2\makebox{Tr}F^4$ effective interactions
of four gauge field strength 
tensors. By using iteration based on the soft gluon limit, we derive a simple formula valid to that order for arbitrary $N$.
We argue that such a procedure can be extended to all orders in
$\alpha'$. If nature gracefully picked a sufficiently low string mass
scale, our results would
be important for studying string effects in multi-jet production at the Large Hadron
Collider (LHC).

\end{abstract}

\maketitle
\section{\label{intro}Introduction and review}
Since Rutherford's times, elementary particle physics relies on scattering experiments.
The physical cross sections, determined by the scattering amplitudes, reflect the properties 
of underlying interactions. In the framework of the standard model, high energy scattering 
experiments allow probing 
inside hadrons, into the gauge interactions of quarks and gluons. Due to the asymptotic 
freedom of Quantum Chromodynamics (QCD), the corresponding amplitudes can be computed 
perturbatively, order by order in the QCD coupling constant. Already at the tree level, 
such computations can be quite complicated, especially when a large number of external 
particles is involved, like in the scattering processes describing multi-jet production 
at hadron colliders.
After more than thirty years of steady progress in perturbative QCD, 
we have a very good understanding of the tree-level scattering,
completely sufficient for studying jet physics in the upcoming Large Hadron Collider (LHC) 
experiments at CERN. Hopefully, LHC will reach beyond the standard model, and the signals of 
new physics will rise above the QCD background.

Although many scenarios have been proposed beyond the standard model, there is no clear 
prediction for the energy scale of new physics. Even if no spectacular effect like, say, a 
direct production of Kaluza-Klein particles, is discovered at LHC, some sub-threshold effects could be observed, due to the presence of contact interactions induced by virtual particles too heavy to 
be produced on-shell. In this paper, we investigate such effects  in superstring-based scenarios, where the scale of new physics is determined by the Regge slope $\alpha'$ of mass dimension ${-}2$. Massless gauge bosons are separated by a mass gap of  $1/\sqrt{\alpha'}$ from the massive string modes. Traditionally, the 
Regge slope and the respective string mass scale had been tied to the Planck mass, however more 
recently, some serious consideration was given to D-brane 
models with much lower string mass scale, possibly even within the reach of LHC 
\cite{Lykken:1996fj,Antoniadis:1998ig}. The full-fledged string amplitudes depend on $\alpha'$,
resulting in large corrections to Yang-Mills (YM) amplitudes once some some kinematic invariants 
characterizing energy scales involved in the scattering process become comparable to $1/\sqrt{\alpha'}$.

We work in the framework of open type I superstring theory compactified to four 
dimensions, with gluons being open string excitations. In the tree approximation, 
the multi-gluon amplitudes 
are computed on a disk worldsheet, with the vertices inserted at the boundary. They do not depend 
on the compactification manifold because they are completely determined by two-dimensional 
superconformal field theory describing four space-time string coordinates of the worldsheet. 
In particular, it does not matter whether supersymmetry is broken or not by compactification. 
Note that the $\alpha'\to 0$ limit is completely determined by pure Yang-Mills theory.

An important feature of open string (disk) computations is that they yield
gluon amplitudes in a very 
particular, 
{\it color-decomposed}\ form:
\be \label{color}
 {\cal A}^{\makebox{\tiny disk}}(\,\{k_i,\lambda_i,a_i\}\,)
 = g^{N-2} \hskip-1.3mm \sum_{\sigma \in S_N/\mathbb{Z}_N} \hskip-1.3mm  
    \makebox{Tr} \, (\, T^{a_{\sigma(1)}}\cdots T^{a_{\sigma(N)}})\ 
     A(\sigma(1^{\lambda_1}),\ldots,\sigma(N^{\lambda_N}))\ ,
\ee
where $g$ is the gauge coupling (${g^2\over4\pi}=\alpha_s$), 
$k_i, \lambda_i$ are the gluon momenta and helicities, and $T^{a_i}$ are matrices in 
the fundamental representation of the gauge group \cite{ts}, describing the color 
states of $N$ gluons. We consider amplitudes with all momenta directed {\it inward}.
Each color trace factor is associated by Chan-Paton rules to one {\it partial amplitude} 
$A(\sigma(1^{\lambda_1}),\ldots,\sigma(N^{\lambda_N}))$ containing all the kinematical 
information.
$S_N$ is the set of all permutations of $N$ objects, while $\mathbb{Z}_N$ is
the subset of cyclic permutations, which preserves the trace;
one sums over the set $S_N/\mathbb{Z}_N$ in order to include all orderings of gluon vertices, 
sweeping out all distinct cyclic orderings in the trace. A similar color decomposition is 
routinely used for multi-gluon amplitudes in QCD 
\cite{Mangano:1990by,Dixon:1996wi}.

In QCD, there exists a subclass of amplitudes that are described, at the tree-level, by
a simple analytic formula valid for arbitrary number $N$ of gluons. Assume that two gluons, with 
the momenta
$k_1$ and $k_2$, in the color states described 
by the matrices $T^{a_1}$ and $T^{a_2}$, respectively, carry negative 
helicities while the rest of gluons, with the momenta and color charges 
$(k_3,T^{a_3}),\dots,(k_N,T^{a_N})$, respectively,
carry positive helicities.
Then the partial amplitude for such a ``maximally helicity violating'' (MHV) configuration,
associated to the $\makebox{Tr} \, (\, T^{a_1}\cdots T^{a_N})$ Chan-Paton factor, is given by
\cite{Parke:1986gb,Berends:1987me}
\begin{equation}
A_{Y\! M}(1^-,2^-,3^+,\dots ,N^+) ~=~ i\ 
\frac{\langle 12\rangle^4}{\langle 12\rangle\langle 23\rangle\langle 34\rangle\cdots\langle N1\rangle}~\equiv~
\mathfrak{M}_{Y\!M}^{(N)}\ .
\label{pt}\end{equation}
%(1^-2^-3^+\dots N^+)
where  $\langle ij\rangle$ are the standard spinor products associated to the momenta 
$k_i$, $k_j$, in the notation of
\cite{Mangano:1990by,Dixon:1996wi}. Other partial amplitudes can be obtained from Eq.(\ref{pt}) 
by applying appropriate permutations to the cyclic denominator $\langle 12\rangle\langle 23
\rangle\langle 34\rangle\cdots\langle N1\rangle$. {}For example,
\begin{equation}\label{ptperm}
A_{Y\! M}(1^-,3^+,2^-,\dots ,N^+) ~=~ i\ 
\frac{\langle 12\rangle^4}{\langle 13\rangle\langle 32\rangle\langle 24
\rangle\cdots\langle N1\rangle}\ .
\end{equation}
Thus the full MHV amplitude is 
\be \label{ptfull}
 {\cal A}_{Y\! M}^{\makebox{\tiny tree}}(1^-,2^-,3^+,\dots, N^+)
 =i \,g^{N-2}\langle 12\rangle^4 \hskip-1.3mm 
\sum_{\sigma \in S_N/\mathbb{Z}_N} \hskip-1.3mm  
   \frac{ \makebox{Tr} \, 
(\, T^{a_{1_{\sigma}}}\cdots T^{a_{N_{\sigma}}})}{\langle 1_{\sigma}2_{\sigma}
\rangle\langle 2_{\sigma}3_{\sigma}\rangle\langle 3_{\sigma}4_{\sigma}
\rangle\cdots\langle N_{\sigma}1_{\sigma}\rangle }\ ,
\ee
where $i_{\sigma}\equiv\sigma(i)$.
The origin of the striking simplicity of MHV amplitudes is not clear. Most likely, it is 
related to some (partial) integrability properties of QCD. There is also an interesting
duality between Yang-Mills theory and twistor strings \cite{Witten:2003nn} that led to a 
new interpretation of Eq.(\ref{pt}) and turned out very useful for developing more efficient 
computational techniques in perturbative QCD \cite{Dixon:2005sp,Cachazo:2005ga}. 
The amplitudes describing non-MHV helicity configurations 
are known to be more complicated. Thus when studying multi-gluon scattering in string theory, 
it is natural to use MHV configurations as a starting point, in order to find out if the 
full-fledged string amplitudes can also be described by some simple analytic formulas valid to 
all orders in $\alpha'$. Our results show that this is indeed the case.

In order to write down the amplitudes in a concise way, it is convenient to introduce the following
notation for the kinematic invariants characterizing $N$-particle scattering: 
\begin{eqnarray}\label{in}
\bl i\br_n&=&\alpha'\ (k_i+k_{i+1}+\dots+k_{i+n})^2\ ,\\[1mm]
&&\hskip -1.4cm\epsilon(i,j,m,n)\;=\, \alpha'^{\, 2}\,\epsilon_{\alpha\beta\mu\nu}\,
k_i^{\alpha}k_j^{\beta}k_m^{\mu}k_n^{\nu}\,  \label{pseudo}\end{eqnarray}
where $k_i$ denotes the momentum of $i$-th particle, with the cyclic identification
$i+N\equiv i$, and $\epsilon_{\alpha\beta\mu\nu}$ is the four-dimensional Levi-Civita symbol.
The factors of $\alpha'$ render the above invariants dimensionless.
Note that the momenta are always subject to the momentum conservation constraint,
$\sum\limits_{i=1}^{N}k_i=0$, and all gluons are on-shell, $k_i^2=\bl i\br_0=0$.
It is also convenient to introduce
\be 
s_{ij}=2\ap \,k_ik_j\ . 
\label{KINI}
\ee
By using momentum conservation, these scalar products can be always expressed in terms of 
$N(N-3)/2$ invariants \req{in}. This is done for $N\le 6$ in Appendix A. Note however, that for $N\ge 6$, the number of independent invariants of 
type  \req{in} is smaller than $N(N-3)/2$, as explained in Section V.

The amplitude for four-gluon scattering has been known for a long time \cite{GS,Schwarz,VectorT}.
All string effects are summarized in one Beta function (Veneziano amplitude)
\be \label{v4} V^{(4)}(k_1,k_2,k_3,k_4)= V^{(4)}(s_1,s_2)=
\frac{\Gamma(1+s_1)\ \Gamma(1+s_2)}{ \Gamma(1+s_1+s_2)}\ ,\ee
where $s_1\equiv\bl 1\br_1=s_{12}$, $s_2\equiv\bl 2\br_1=s_{23}$,
as the formfactor of Yang-Mills amplitude: \begin{equation}\label{m4}
A(1^-,2^-,3^+ ,4^+) =V^{(4)}(s_1,s_2)\ \mathfrak{M}_{Y\!M}^{(4)}\ ,
\end{equation}
with $ \mathfrak{M}_{Y\!M}^{(N)}$ defined in Eq.(\ref{pt}). An obvious but very important property of the Veneziano formfactor $V^{(4)}(k_1,k_2,k_3,k_4)$ is its invariance under the cyclic permutations of the momenta. All other partial amplitudes can be obtained by applying the coset permutations $\sigma$ to the right hand side of
Eq.(\ref{m4}), \cf Eq.(\ref{ptfull}), now including also the cyclic formfactor. Thus the full four-point amplitude is
\be \label{stfull}
 {\cal A}^{\makebox{\tiny disk}}(1^-,2^-,3^+,4^+)
 =i \,g^{2}\langle 12\rangle^4 \hskip-1.3mm \sum_{\sigma \in S_4/\mathbb{Z}_4} \hskip-1.3mm  
   \frac{ \makebox{Tr} \, (\, T^{a_{1_{\sigma}}}T^{a_{2_{\sigma}}}T^{a_{3_{\sigma}}}T^{a_{4_{\sigma}}})\ 
V^{(4)}(k_{1_{\sigma}},k_{2_{\sigma}},k_{3_{\sigma}},k_{4_{\sigma}})}{\langle 1_{\sigma}2_{\sigma}
\rangle\langle 2_{\sigma}3_{\sigma}\rangle\langle 3_{\sigma}4_{\sigma}\rangle
\langle 4_{\sigma}1_{\sigma}\rangle }\ .
\ee 
In fact, for four (and five) gluons, all non-MHV amplitudes are vanishing \cite{Parke:1986gb,Parke:1985pn}, 
therefore the above expression captures the full amplitude.
It can be expanded in powers of $\alpha'$ by using
\be\label{v4exp} V^{(4)}(s_1,s_2)\approx 1-\frac{\pi^2}{6}\ 
s_1s_2+\zeta(3)\  s_1s_2\ (s_1+s_2)+{\cal O}(\alpha'^{4})\, .\ee
The leading string correction to the Yang-Mills amplitude, which originates from the second term in the above expansion, of order ${\cal O}(\alpha'^2)$,
has been extensively discussed in the literature \cite{GS,Schwarz,VectorT}. It is due to the 
following contact interaction term of four gauge field strength tensors:
\begin{eqnarray}\label{if4}
{\cal I}_{F^4} &=&-\frac{\alpha'^2\pi^2}{6}\ \makebox{Tr}\big(\,  {F_{\mu_1\mu_2}}
{F_{\mu_2\mu_3}}
{F_{\mu_3\mu_4}}{F_{\mu_4\mu_1}} + 2\,{F_{\mu_1\mu_2}}
{F_{\mu_3\mu_4}}{F_{\mu_2\mu_3}}
{F_{\mu_4\mu_1}}\nonumber\\ &&\qquad\qquad
-\frac{1}{4}\,{F_{\mu_1\mu_2}}
{F_{\nu_1\nu_2}}{F_{\mu_2\mu_1}}
{F_{\nu_2\nu_1}}\, -\frac{1}{2}\,{F_{\mu_1\mu_2}}{F_{\mu_2\mu_1}}
{F_{\nu_1\nu_2}}
{F_{\nu_2\nu_1}}\big)\ ,\end{eqnarray}
where the color trace is taken with the  tensors $F_{\mu\nu}$ in the fundamental 
representation. This interaction term will play an important role in the further 
discussion of $N$-point amplitudes. Formally, it can be obtained from the  
${\cal O}(\alpha'^2F^4)$ term appearing in the low-energy expansion of the 
Born-Infeld Lagrangian of 
non-linear electrodynamics, by applying to it Tseytlin's ``symmetrized trace'' 
prescription \cite{Tseytlin:1999dj}. Note that for Abelian gauge bosons, the 
pure Yang-Mills part of the amplitude (\ref{stfull}) cancels after summing over all 
coset permutations and the low-energy expansion begins with the Born-Infeld contribution.

The paper is organized as follows. In Section II, we give a brief description of the 
formalism used for calculating multi-gluon amplitudes on a disk world-sheet. Integrations
over the vertex positions yield certain generalized hypergeometric functions, 
their number increasing dramatically with the number of gluons, 
therefore in addition to handling a cumbersome algebra, one has to figure out how to 
construct a proper basis of the boundary integrals. In Section III, we rewrite the 
five-gluon amplitude in an MHV form similar to four gluons, {\it c.f}.\ Eq.(\ref{m4}), 
in terms of two independent (hypergeometric) functions of kinematic invariants. One of 
them plays the role of the Veneziano formfactor, while the second is associated to 
the ${\cal I}_{F^4}$ contact term (\ref{if4}). In Section IV, we extract the MHV part 
of the six-gluon amplitude. Here, all kinematic information is contained in six 
``triple'' generalized hypergeometric functions. We discuss
the low-energy behavior of the amplitude and check that it satisfies all constraints based on
permutation symmetries and soft/collinear limits. We 
show that it is possible to reconstruct the result, obtained from tedious calculations, by imposing these constraints on a general ansatz.
In Section V, we proceed to the general $N$-gluon case. 
We show that all ${\cal O}(\alpha'^2)$ order string corrections to MHV Yang-Mills
amplitudes originate from the interactions associated to the ${\cal I}_{F^4}$ effective action term. 
By iteration, we obtain a simple $N$-gluon formula valid to that order and outline
a recursive procedure that could make possible a complete determination
of all MHV amplitudes, to all orders in $\alpha'$. In Conclusions, we discuss our results in 
a broader context of QCD and superstring theory. The paper contains three appendices. In Appendix A we summarize some aspects of the scattering kinematics for $N=4,5$ and 6 gluons.
In Appendices B and C we discuss various properties of the generalized (triple) hypergeometric functions describing the six-gluon amplitude.

Some results of this work have been already reported in our 
Letter \cite{Stieberger:2006bh}.

\section{\label{multidisk}Multi--gluon scattering on the disk}

In this Section, we review the general structure of multi-gluon string amplitudes, focusing 
on the computational problems related to a large number of vertices at the boundary. 
It can be skipped by readers who are not interested in technical details.

We are interested in superstring theory with gluons coming from open strings. A variety of 
four-dimensional models can be constructed, each of them described by a two-dimensional
superconformal field theory (SCFT). At the disk (tree) level, the details of the ``internal'' 
part of SCFT
associated to the compactification space do not affect the scattering amplitudes of 
four-dimensional gauge bosons. Furthermore, the entire disk boundary is attached to a 
single stack of D-branes. Thus without losing generality, 
we can consider type I theory with D9-branes and sixteen supercharges.
Nevertheless, our discussion holds for both type I or type IIA/B
theories with Dp-branes and any gauge group. Spacetime supersymmetry can be preserved or 
broken by the internal
space or by D-brane configurations.

Gluons originate from the  excitations of string space-time
coordinates $X^\mu$ and their SCFT partners $\psi^\mu$, satisfying Neumann boundary 
conditions on the world-sheet. From all other SCFT fields, only the reparametrization 
ghost $c$ and the scalar $\phi$ bosonizing the superghost system will enter explicitly 
into our computations. In the $(-1)$-ghost picture, the vertex operator 
for a gluon with momentum $k$, polarization $\xi$ (or helicity $\lambda$)
and color state $a$ is given by
\be\label{vertex}
V^{(-1)}(z,\{k,\xi,a\})=T^a\ \xi_\mu\ e^{-\phi(z)}\ \psi^\mu(z)\ e^{ik_\rho X^\rho(z)}\ ,
\ee
where $z$ is the vertex position at the disk boundary. Note that the color state $a$ 
is represented by the matrix $T^a$ in the fundamental representation of the gauge group.
In the zero-ghost picture, this vertex operator is given by:
\be\label{vertexi}
V^{(0)}(z,\{k,\xi,a\})=T^a\ \xi_\mu\ [\ \p X^\mu(z)+i\ (k\psi)\ \psi^\mu(z)\ ]\
e^{ik_\rho X^\rho(z)}\ .
\ee

The disk may be conformally mapped to the upper half plane $\im z\geq 0$ 
with the real axis as its boundary. Hence all vertex positions are located
on the real axis.
The $N$-gluon disk amplitude is:
\begin{eqnarray}\nonumber
{\cal A}^{\makebox{\tiny disk}}(\,\{k_i,\xi_i,a_i\}\,)
&=& 
\sum_{\sigma \in S_N/\mathbb{Z}_N} \hskip-1.3mm  
    \makebox{Tr} \, (\, T^{a_{1_{\sigma}}}\cdots T^{a_{N_{\sigma}}})\ V_{\rm CKG}^{-1}\ 
\int_{-\infty}^{\infty} dz_{1_{\sigma}} \int_{z_{1_{\sigma}}}^{\infty} dz_{2_{\sigma}}
\dots\int_{z_{(N{-}1)_{\sigma}}}^{\infty} dz_{N_{\sigma}}\\ \label{START}&&
\hskip 2.5cm\times~\vev{V^{(-1)}(z_1)\ V^{(-1)}(z_2)\ V^{(0)}(z_3)\dots
V^{(0)}(z_N)}\, ,
\end{eqnarray}
where the color part of the vertices has been factored out by following the Chan-Paton rule.
In the above expression, $V_{\rm CKG}$ is the volume of the conformal
Killing group $PSL(2,\mathbb{R})$ which leaves the boundary [$\im(z)=0$] of the disk fixed.
It will be canceled by fixing three
positions and introducing the respective $c$-ghost correlator.
Note that two vertices are inserted in the $(-1)$-ghost picture in
order to cancel the background ghost charge. 

By comparing Eq.\req{START} with the color-decomposed form of ${\cal
  A}^{\makebox{\tiny disk}}$, 
see Eq.\req{color}, we see that the partial amplitude
$A(\sigma(1^{\lambda_1}),\ldots,
\sigma(N^{\lambda_N}))$ 
is obtained by integrating the correlator of the vertex operators over the region
$\{\ -\infty< z_{1_{\sigma}}<z_{2_{\sigma}}<\ldots<z_{N_{\sigma}}<\infty\ \}$.
In the following, we shall concentrate on the Chan-Paton factor
$\Tr(\, T^{a_{1}}\cdots T^{a_{N}})$, \ie in Eq.\req{START}
we pick up the integration region ${\cal R}\equiv\{-\infty<  z_{1}<z_{2}<\ldots<z_{N}<\infty\ \}$
and compute
\be\label{compute}
A(1^{\lambda_1},\ldots,N^{\lambda_N})=
V_{\rm CKG}^{-1}\ \int_{\cal R}\Big(\prod_{r=1}^N\ d z_r\Big)\,
\vev{V^{(-1)}(z_1)\ V^{(-1)}(z_2)\ V^{(0)}(z_3)\dots
V^{(0)}(z_N)}\ .
\ee
Due to the $PSL(2,\mathbb{R})$ invariance on the disk, we can 
fix three positions of the vertex operators. A convenient choice is
\be\label{Choice}
z_1=-z_\infty=-\infty\ \ \ ,\ \ \ z_2=0\ \ \ ,\ \ \ z_3=1\ ,
\ee
which implies the ghost factor $\vev{c(z_1)c(z_2)c(z_3)}=-z_\infty^2$.
The remaining $N-3$ vertex positions $z_4,\ldots,z_N$
take arbitrary values inside the integration domain $\cal R$. 
It is convenient to use the following parameterization:
\be\label{convenient}
z_4=x_1^{-1}\ \ \ ,\ \ \ z_5=(x_1x_2)^{-1}\ \ \ ,\ \ \ z_6=(x_1x_2x_3)^{-1}\ \ \ ,\ \ \ 
\ldots\ \ \ ,\ \ \ z_N=\prod_{i=1}^{N-3} x_i^{-1}\ ,
\ee
with $0<x_i<1$. The corresponding Jacobian is  $\big| \partial z_i/\partial x_j\big|=
\prod\limits_{r=1}^{N-3} x_r^{1+r-N}$.

The correlator of vertex operators in Eq.\req{START} is evaluated by
performing all possible 
Wick contractions. It decomposes into products of two-point
functions, introducing kinematic factors consisting of the scalar
products of momentum and 
polarization vectors, of the form $k_ik_j,\ \xi_ik_j$ and $\xi_i\xi_j$.
Schematically, one obtains
\be\label{tothisend}
A(1^{\lambda_1},\ldots,N^{\lambda_N})=\sum_I
 \Kc_I\ F\lf[{n^I_a \atop n^I_{ab}}\ri]\ ,
\ee
where each $\Kc_I$ consists of products of such kinematic factors while the
respective integrals can 
be written as
\begin{eqnarray}
{\textstyle F\lf[{n_a \atop n_{ab}}\ri]}&\equiv& \int_0^1 dx_1\ldots \int_0^1 dx_{N-3} 
\prod_{a=1}^{N-3}x_a^{1+a-N+n_a}\ \prod_{b=a}^{N-3} x_a^{2\alpha'\,k_{b+3}
\lf(k_1+\sum\limits_{j=a+3}^{b+2} k_j\ri)}\nonumber\\ &&\hskip 6cm \times
\lf(1-\prod_{j=a}^b x_j\ri)^{2\alpha'\,k_{2+a}k_{3+b}+n_{ab}}
\label{FULL}
\end{eqnarray}
with the indices $b\ge a=1,2,\dots,N{-}3$, and the integers $n_a,\,
n_{ab}$ taking values 
$0,~\pm 1$ or $\pm 2$. By convention, the sum in the exponent is zero
for $b=a$. The integral 
involves $N(N-3)/2$ different Laurent polynomials in $x_a$. Their integer
powers $n_a,\, n_{ab}$ control the
physical poles of the amplitude, 
in $N(N-3)/2$ invariant masses of dual resonance channels involving
$2,3,\dots,E(\frac{N}{2})$ 
external particles ($E$ denotes the integer part).  

{}For $N=4$, the integral \req{FULL} yields the Beta function
\be\label{boilo}
{\textstyle F\lf[{n_1\atop n_{11}}\ri]}=\int_0^1 dx_1\ x_1^{-2+s_{23}+n_1}\ 
(1-x_1)^{s_{12}+n_{11}}=\fc{\F{2}{1}\lf[{s_{23}+n_1-1\ ,\ 
-s_{12}-n_{11}\atop s_{23}+n_1}\ ;\ 1\ri]}{s_{23}+n_1-1}\ \ .
\ee
{}For $N=5$, one obtains the hypergeometric function $\F{3}{2}$ 
\cite{Medina:2002nk}:
\begin{eqnarray}\nonumber
\hskip-0.5cm {\textstyle F\lf[{n_1,n_2\atop n_{11},n_{12},n_{22}}\ri]}&=&
\int\limits_0^1 dx_1\int\limits_0^1 dx_2\ 
x_1^{-3+s_{23}+n_1}\  x_2^{-2+s_{15}+n_2}\\[0mm] \nonumber
&&\hskip 2cm \times \,(1-x_1)^{s_{34}+n_{11}}\ 
(1-x_2)^{s_{45}+n_{22}}\ (1-x_1x_2)^{s_{35}+n_{12}}\\[3mm]
&&\label{boil}\hskip -3cm =\fc{\Gamma(s_{23}+n_1-2)\ \Gamma(s_{15}+n_2-1)\ \Gamma(s_{34}+n_{11}+1)
\ \Gamma(s_{45}+n_{22}+1)}{\Gamma(s_{23}+s_{34}+n_1+n_{11}-1)\ 
\Gamma(s_{15}+s_{45}+n_2+n_{22})}
\\[2mm] \nonumber && \textstyle \hskip 2cm\times \,
\F{3}{2}\lf[{s_{23}+n_1-2\ ,\ s_{15}+n_2-1\ ,\ 
-s_{35}-n_{12}\atop  s_{23}+s_{34}+n_1+n_{11}-1\ ,\ 
s_{15}+s_{45}+n_2+n_{22}}\ ;\ 1\ri]\ .
\end{eqnarray}
Both integrals \req{boilo} and \req{boil} boil down to hypergeometric
functions of one variable, \ie some $\F{p}{q}\lf[{a_1,\ldots,a_p\atop b_1,\ldots,b_q}; u=1\ri]$.
However, this pattern does not persist beyond $N=5$, due to the form
of the integrand \req{FULL} that does not fit into any hypergeometric function of
one variable~$u$.
%Starting at $N=6$ there is  the set of $\h(N-5)(N-2)$ additional  polynomials 
%\be\label{newpoly}
%\lf\{\ \lf(1-\prod_{j=l}^ix_j\ri)\ \ \ |\ \ l<i\leq N-3\ \ri\}\ee
%for $l\neq 1 \wedge i\neq N-3$ \cite{count}.
%{\it E.g.} these polynomials are
%\begin{eqnarray}
%N=6&:&\{\ (1-x_1x_2),\ (1-x_2x_3)\ \}\\ \nonumber
%N=7&:&\{\ (1-x_1x_2),\ (1-x_2x_3),\ (1-x_3x_4),\ (1-x_1x_2x_3),\
%(1-x_2x_3x_4)\ \}\ .
%\end{eqnarray}
%for $N=6$ and $N=7$, respectively.
%With those polynomials the integrals \req{FULL} do not anymore represent 
%hypergeometric functions (of one variable), \ie any $\F{p}{q}[x]$, but
In general, one obtains multiple Gaussian hypergeometric series, more precisely 
certain generalized Kamp\'e de F\'eriet functions \cite{srivastava}. 
For example, for $N=6$, the integral
% we obtain the triple hypergeometric function 
%$F^{(3)}$~\cite{Oprisa:2005wu}:
\begin{eqnarray}\label{triple} 
{\textstyle F\lf[{n_1,n_2,n_3\atop n_{11},n_{12},n_{22},n_{13},n_{23},n_{33}}\ri]}&=&
\int\limits_0^1 \! dx_1\!\int\limits_0^1\! dx_2\! \int\limits_0^1\! dx_3\ 
x_1^{-4+s_{23}+n_1}\  x_2^{-3+\ap (k_2+k_3+k_4)^2+n_2}\ 
x_3^{-2+s_{16}+n_3} \nonumber  \\[1mm] 
&&\quad\times\, (1-x_1)^{s_{34}+n_{11}}\ (1-x_2)^{s_{45}+n_{22}}\ 
(1-x_3)^{s_{56}+n_{33}} \\[1mm] \nonumber
&&\quad\times\, (1-x_1x_2)^{s_{35}+n_{12}} 
\ (1-x_2x_3)^{s_{46}+n_{23}}\ (1-x_1x_2x_3)^{s_{36}+n_{13}}\, 
\end{eqnarray}
can be expressed in terms of the triple hypergeometric function $F^{(3)}$~\cite{Oprisa:2005wu}.

A very important part of the discussion of scattering amplitudes is
the examination of their low-energy behavior. To that end, the integrals \req{FULL} must be
expanded in powers of $\ap$. One can first expand the integrand and 
then integrate the series term after term.
A typical, but by far not the most general, class of integrals  that appear in this way are:
\be
\zeta(s_1,\ldots,s_k)=\lf(\prod_{j=1}^k \fc{(-1)^{s_j-1}}{\Gamma(s_j)}\ri)
\int_0^1 dx_1\ldots \int_0^1 dx_k \ \ 
\prod_{j=1}^{k} \  x_j^{k-j}\ \fc{(\ln x)^{s_j-1}}{1-\prod\limits_{i=1}^jx_i}\ .
\label{Polyintegral}
\ee
They integrate to multiple zeta values of length $k$ \cite{bbbl}:
\be\label{genzeta}
\zeta(s_1,\ldots,s_k)=\sum_{n_1>\ldots>n_k>0}\ \prod_{j=1}^k\ \fc{1}{n_j^{s_j}}=
\sum_{n_1,\ldots,n_k=1}^\infty\ \prod_{j=1}^k\ \lf(\sum\limits_{i=j}^k n_i\ri)^{-s_j}\ ,
\ee
with $s_1\geq 2\ ,\ s_2,\ldots,s_k\geq 1$.
Such integer series are completely sufficient for discussing the expansions of amplitudes
involving four and five gluons however, as mentioned before,
starting at $N=6$,  more general classes of integrals appear.
Their expansions involve multiple harmonic series and  generalized Euler-Zagier sums.
We refer interested readers to
Ref.\cite{Oprisa:2005wu} for a detailed account on the relation between 
multiple Gaussian hypergeometric functions and  Euler--Zagier sums.
Actually, the integer sums that appear 
in the context of multi-gluon string scattering play an important role in modern number theory 
\cite{zagier}.

The number of independent (with respect to the momentum conservation constraint) kinematic 
factors $\Kc_I$ and of the
associated functions $F\lf[{n_a^I\atop n_{ab}^I}\ri]$ entering into the $N$-gluon partial 
amplitude \req{tothisend} grows with $N$. In our analysis, we
encountered $77$ functions for $N=5$ and $1,270$ functions for $N=6$, although these numbers 
may vary depending on the implementation of  
momentum conservation constraints {\it etc}. In fact, many functions are related by means of 
polynomial relations of their integrands or by partial integration. The only systematic way 
of handling them for arbitrary $N$ is to find a basis, consisting of an {\it a priori\/} 
unknown number $\nu_N$ of functions, and to express all other functions
as linear combinations of the
basis elements with the coefficients given by some rational (homogeneous) functions of the 
kinematic invariants~\req{in}.
This program has been successfully implemented   in 
\cite{Medina:2002nk,Barreiro:2005hv} for $N=5$ and in \cite{Oprisa:2005wu} for $N=6$ and 
will be continued in \cite{prep}.
For a given $N$, an efficient way of generating systems of equations relating the integrals 
\req{FULL},
that can be used to find a minimal set of independent functions, is based on world-sheet 
supersymmetry \cite{Oprisa:2005wu}. It works in the following way.
In Eq.\req{START}, the two vertices in the $(-1)$-ghost picture were inserted, for convenience, 
at $z_1$ and $z_2$. However, due to world-sheet supersymmetry, they could be inserted at any 
other two points, hence there are $\lf(N \atop 2\ri)$ ways of computing the same amplitude 
that should give the same answer for the coefficients of all (independent) kinematic factors 
$\Kc_I$ in Eq.\req{tothisend}.
By comparing these coefficients, one obtains many relations among the integrals \req{FULL}.
The corresponding set of equations is always under-determined and may be solved by 
expressing all functions \req{FULL} in terms of a $\nu_N$-dimensional  basis.
Of course, the dimension of the
space of functions grows
with the number of  gluons: $\nu_4=1,\ \nu_5=2,\ \nu_6=~6,~\ldots$

Although only one partial amplitude $A(1^{\lambda_1},\ldots,N^{\lambda_N})$ has been discussed 
here explicitly, all other partial amplitudes 
$A(\sigma(1^{\lambda_1}),\ldots,\sigma(N^{\lambda_N}))$ can be obtained in exactly the same way.
As we shall see in the following, a convenient choice of $\nu_N$ basis functions is dictated
by various physical properties of the amplitude \req{color}.

\section{\label{five}Five Gluons}

The purpose of this section is to summarize the results of five-gluon computations 
\cite{Medina:2002nk,Barreiro:2005hv}
and to rewrite the five-gluon amplitude in the four-dimensional helicity basis. Recall that
up to five gluons, the  amplitudes are purely MHV.
Here, five invariants
are necessary to specify the kinematics. They can be chosen as
$s_i\equiv\bl i\br_1,~i=1,\dots,5$, {\it i.e}.\ as the  cyclic  orbit of
$\bl 1\br_1$ obtained by the action of $\mathbb{Z}_5$ subgroup of cyclic permutations, generated by $i\to i+1$ mod 5 \cite{appa}.
%In addition, all Levi-Civita pseudoscalars can be expressed in terms of %$\epsilon_1\equiv\epsilon(1,2,3,4)$
%by using the momentum conservation law.

The integrals over two vertex positions have the form \req{boil}, specified by
five integers $n_1,n_2,n_{11},n_{12},n_{22}$.
One finds  \cite{Barreiro:2005hv,Oprisa:2005wu} that all integrals can be expressed 
in terms of just two functions:
\be\label{onefinds}\ts
 f_1=F\lf[{2,1\atop 0,0,0}\ri]\qquad\makebox{and}\qquad f_2=F\lf[{3,2 \atop 0,-1,0}\ri]
\ .\ee
By means of simple algebraic operations and partial integrations it is easy to see that these functions transform in the following way under the $\mathbb{Z}_5$ generator $i\to i+1$ mod 5:
\begin{eqnarray}
f_1&\to&
{\ts F\lf[{4,2\atop -1,-1,0}\ri]}
=\fc{1}{s_1s_3}\left[\ s_2s_5\ f_1+(s_2 s_3 - s_3 s_4 - s_1 s_5 + s_4
  s_5)\ f_2\ \right]\, ,\\ \ts
f_2&\to& f_2\, .
\label{cyc5}
\end{eqnarray}

In the notation of \cite{Barreiro:2005hv}, the partial amplitude
\be\label{ampl5}
A(1^{\lambda_1},2^{\lambda_2},3^{\lambda_3},4^{\lambda_4},5^{\lambda_5})=T\cdot A_{Y\!M}(1^{\lambda_1},2^{\lambda_2},3^{\lambda_3},4^{\lambda_4},5^{\lambda_5})+K_3\cdot A_{F^4}(1^{\lambda_1},2^{\lambda_2},3^{\lambda_3},4^{\lambda_4},5^{\lambda_5})\, \ee
where
\be T(s_i)=s_2 s_5\ f_1 + (s_2 s_3 + s_4 s_5)\ f_2\,\qquad\makebox{and}\qquad K_3(s_i)=f_2\, .\ee
In Eq.\req{ampl5}$, A_{Y\!M}$ is the tree-level Yang-Mills amplitude while $A_{F^4}$
is generated by the ${\cal I}_{F^4}$ interaction term \req{if4} discussed in the Introduction.
\begin{figure}
\includegraphics[scale=0.7]{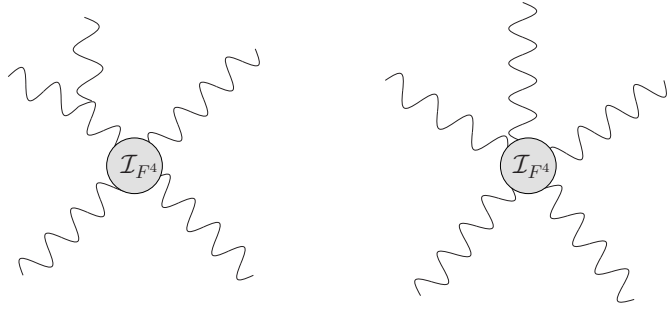}
\caption{\label{fig:mhvfig1} Feynman diagrams contributing to the $A_{F^4}$ part of the five-gluon amplitude involve a single four- or five-gluon vertex due to the ${\cal I}_{F^4}$ effective interaction, represented here by the blob.}
\end{figure}
%The amplitudes $A_{Y\!M}$ and $A_{F^4}$ consist of many Lorentz-invariant terms falling %into two classes: with one or two factors of $\xi(i)\cdot\xi(j)$ contractions between the %polarization vectors \cite{polar}. The amplitude
%$A_{F^4}$ has the following field-theoretical interpretation.
There are two Feynman diagrams, shown in Figure 1, that combine to $A_{F^4}$: the diagram with ${\cal I}_{F^4}$ four-gluon vertex
including one off-shell gluon decaying into two external gluons via the standard three-gluon
Yang-Mills interaction, and the diagram with
${\cal I}_{F^4}$ five-gluon vertex. The function $K_3$ in Eq.(\ref{ampl5}) can then be interpreted as
a string ``formfactor'' of $\alpha'^{\, 2}{\rm Tr}\, F^4$ interactions, playing role similar to the Yang-Mills formfactor $T$.

The amplitudes  $A_{Y\!M}$ and $A_{F^4}$, as well as the functions $T(s_i)$ and $K_3(s_i)$ are invariant under 
cyclic permutations, therefore the amplitude (\ref{ampl5}) is cyclic invariant. Furthermore, it has the correct factorization properties into four-gluon amplitudes,
in the limit of zero (soft) momentum of one gluon and in the limit of two parallel momenta \cite{Barreiro:2005hv}. The low-energy behavior of the amplitude is determined,
up to the order ${\cal O}(\alpha'^3)$, by the following expansions:
\begin{eqnarray}\nonumber
f_1&=&\fc{1}{s_2s_5}-\fc{\pi^2}{6}\ \lf(\fc{s_4}{s_2}+\fc{s_3}{s_5}\ri)\\ 
&&+~ \label{rexp} \zeta(3)\ 
\lf(\ -s_1+s_3+s_4+\fc{s_4^2}{s_2}+\fc{s_2s_3}{s_5}+\fc{s_3^2}{s_5}+
\fc{s_4s_5}{s_2}\ \ri)+\dots\ ,\\
f_2&=&\frac{\pi^2}{6}-\zeta(3)\ (s_1+s_2+s_3+s_4+s_5)+\dots\, .\label{rexp2}
\end{eqnarray}

In order to rewrite the amplitude (\ref{ampl5})  in the MHV form, we evaluate it for the specific configuration of the polarization vectors, choosing a  gauge
with the most convenient  ``reference momenta'' \cite{Mangano:1990by,Dixon:1996wi}. We choose the reference momenta
$k_5$ for $\xi^-(1,2)$ and $k_1$ for $\xi^+(3,4,5)$:  \be  \displaystyle
\xi^-_{\mu}(i)~=~-\frac{\langle k_5^+|\gamma_{\mu}|k_i^+\rangle}{\sqrt{2}\,[5\,i\,]} ~\makebox{for}~i=1,2 \qquad\makebox{and}\qquad
\xi^+_{\mu}(j)~=\quad\frac{\langle k_1^-|\gamma_{\mu}|k_j^-\rangle}{\sqrt{2}\,\langle 1j\,\rangle} ~\makebox{for}~ j=3,4,5.
\label{ximu}\ee
Indeed, with such a choice, the only non-vanishing scalar products
of the polarization vectors are
\be \xi^-(2)\cdot\xi^+(3)=-\frac{\langle 12\rangle [35]}{\langle 13\rangle [25]}\qquad
\makebox{and}\qquad \xi^-(2)\cdot\xi^+(4)=-\frac{\langle 12\rangle [45]}{\langle 14\rangle [25]}\, .
\label{xixi}\ee
In this gauge, the respective kinematic factors $\Kc_I$, see Eq.\req{tothisend}, contain only one
$\xi_i\xi_j$ factor while the remaining three polarization vectors are contracted with the momenta.
The computation consists of manipulations with spinor products, involving a repeated use of the momentum conservation law and of Schouten identity \cite{Mangano:1990by,Dixon:1996wi}. A very useful check is provided by the cancellation of unphysical poles $[5\,i\,]^{-1}$
and $\langle 1j\,\rangle^{-1}$ introduced by the choice (\ref{ximu}) of the reference momenta. After factorizing out the Yang-Mills MHV amplitude $\mathfrak{M}_{Y\!M}^{(5)}$, \cf Eq.\req{pt},
the remaining spinor products can be either expressed in terms
of  kinematic invariants $s_i$ or they form the products \cite{Mangano:1990by,Dixon:1996wi}
\begin{eqnarray}\label{tprod}
\alpha'^2\, T[i,j,l,m]&\equiv& \alpha'^2\,\langle ij\rangle [j\, l]\langle l\,m\rangle[mi] = \alpha'^2 \,\makebox{tr}\left(
\frac{1}{2}(1-\gamma_5) \ksl_i \ksl_j \ksl_l \ksl_m \right) \\
&=& \frac{1}{2}\ \left[\ 
s_{ij}\ s_{lm} - s_{il}\ s_{jm} + s_{im}\ s_{jl}
  - 4i\ \epsilon(i,j,l,m) \ \right]\ . 
\end{eqnarray}
The new feature, as compared to four gluons, is the appearance of Levi-Civita pseudoscalars. They originate from the $A_{F^4}$ part of the amplitude only. {}For five gluons, the momentum conservation law allows expressing all such pseudoscalars in terms of one of them, that can be chosen to be
$\epsilon(1,2,3,4)$. The final result is
\begin{equation}\label{m5}
A(1^-,2^-,3^+,4^+,5^+)=
[\ V^{(5)}(s_j)-2i\ 
P^{(5)}(s_j)\ \epsilon(1,2,3,4)\ ]\ \mathfrak{M}_{Y\!M}^{(5)},\end{equation}
where
\be\label{v5}
V^{(5)}(s_i)= s_2s_5\ f_1 + \frac{1}{2}\ (s_2s_3+s_4s_5-s_1s_2-s_3s_4-s_1s_5)\ f_2
\quad\makebox{and}\quad P^{(5)}(s_i)= f_2 \, .
\ee
The above functions, as well as the pseudoscalar $\epsilon(1,2,3,4)$, are invariant under cyclic permutations \cite{cyc},
thus the 
factor multiplying the Yang-Mills amplitude in Eq.\req{m5} is cyclic invariant.

The low-energy behavior of the amplitude
(\ref{m5}) is determined by the  expansions (\ref{rexp}):
\begin{eqnarray}
\nonumber
V^{(5)}(s_i)&=&  1-\frac{\pi^2}{12}\ \{s_1s_2\}\\  &&+~ \label{v5ex} \frac{\zeta(3)}{2}\  
(\ \{s_1^2s_2\}+\{s_1s_2^2\}+\{s_1s_3s_5\}\ )+\dots\ ,\\
P^{(5)}(s_i)&=& \frac{\pi^2}{6}-\zeta(3)\ \{s_1\}+\dots\ ,\nonumber
\end{eqnarray}
where the curly brackets enclosing kinematic invariants imply the summation over all distinct elements of the respective cyclic orbit \cite{cexpl}.

The connection to the four-gluon amplitude (\ref{m4}) can be established by
considering the  soft limit, say one $k_i\to 0$, see Appendix A. Then the pseudoscalar part of the factor
disappears due to the momentum conservation while the function
\begin{equation}\label{s5}V^{(5)}(s_i)\xrightarrow[k_i= 0]{}
\frac{\Gamma(1+s_1)\ \Gamma(1+s_2)}{ \Gamma(1+s_1+s_2)}
\end{equation}
reproduces the Veneziano formfactor in Eq.(\ref{v4}).

All other partial amplitudes $A(\sigma(1^-),\sigma(2^-),\sigma(3^+),\sigma(4^+),\sigma(5^+))$, which according to Eq.\req{color} are necessary for constructing the full MHV amplitude ${\cal A}^{\makebox{\tiny disk}}(1^-,2^-,3^+,4^+,5^+)$ are obtained from $A(1^-,2^-,3^+,4^+,5^+)$ by simply applying the coset permutations $\sigma$ to the right hand side of Eq.\req{m5}.

\section{\label{six}Six Gluons}

The step from five to six gluons is highly non-trivial. Even in QCD,
the original calculation \cite{Parke:1985ax} used some of the most
advanced tools available at that time,
like extended supersymmetry \cite{Parke:1985pn}, a special choice of
the color factor  basis {\em etc}. 
In addition to algebraic complications due to large numbers 
of Wick contractions and of the associated kinematic factors,
there is a new physics element appearing
at the six-gluon level:
the scattering amplitudes allow also some non-MHV helicity configurations.
Furthermore, each kinematic factor brings an integral
over three vertex positions. The new challenge is to find relations between more than one 
thousand of such integrals and to express them in a suitable basis. Before discussing this problem,
we review the six-particle kinematics (see also Appendix A), which also exhibits some new features as compared to
the five-particle case.

\subsection{\label{kin6}Six-Particle Kinematics}

In five and more dimensions, the number of independent kinematic
invariants in a six-particle 
scattering process can be counted by using the momentum conservation law. There are nine 
invariants that can be grouped into two irreducible representations of
the $\mathbb{Z}_6$ cyclic group generated by  $i\to i+1$ mod 6:
\be s_i\equiv\bl i\br_1,~i=1,\dots,6,\qquad\qquad t_j\equiv\bl j\br_2,~j=1,2,3,\ee
{\it i.e}.\ the $\mathbb{Z}_6$ orbits of $\bl 1\br_1$ and $\bl 1\br_2$ \cite{appa}.
In four dimensions, however,
these variables are subject to a fifth-order polynomial constraint \cite{asribekov}
that reduces the number of independent invariants from nine to eight.
This is due to the trivial fact that in four dimensions,
at most four momentum vectors can be linearly independent. 
Then the columns of the five by five Gram matrix $s$ built of the
elements $s_{ij}$,
% \req{KINI} 
$1\leq i,j\leq 5$, cannot be linearly independent, therefore $\det s=0$.

The vanishing of the Gram determinant is closely related to
the following identity involving the metric
tensor $g_{\mu\nu}$ and the Levi-Civita tensor
$\epsilon_{\alpha\beta\gamma\delta}$:
\be\label{geps} 2g_{\mu\nu}\epsilon_{\alpha\beta\gamma\delta}=g_{\mu\alpha}
\epsilon_{\nu\beta\gamma\delta}+g_{\mu\beta}\epsilon_{\alpha\nu\gamma\delta}
+g_{\mu\gamma}\epsilon_{\alpha\beta\nu\delta}
+g_{\mu\delta}\epsilon_{\alpha\beta\gamma\nu}
+(\mu\leftrightarrow\nu).\ee
%Let us define
%\be\epsilon(i,j,l,m)=\epsilon_{\alpha\beta\gamma\delta}
%\,k_i^{\alpha}
%k_j^{\beta}k_l^{\gamma}k_m^{\delta}\ee
One can eliminate one
four-momentum, say $k_6$, by using momentum conservation, and
define the following pseudoscalars:
\be\epsilon_1=\epsilon(2,3,4,5)
\quad\epsilon_2=\epsilon(1,3,4,5)
\quad\epsilon_3=\epsilon(1,2,4,5)\quad\epsilon_4=\epsilon(1,2,3,5)
\quad\epsilon_5=\epsilon(1,2,3,4)\, ,\label{pseudo6}\ee and further define the
five-component vector
$\epsilon=(\epsilon_1,-\epsilon_2,\epsilon_3,-\epsilon_4,\epsilon_5).$
Then the identity (\ref{geps}) implies
\be
\hskip-0.3cm v\equiv\epsilon\cdot  s=\lf(
\begin{array}{c}
-s_1\ \epsilon_2-(s_1+s_2-t_1)\ \epsilon_3-(s_2+s_5-t_1-t_2)\
\epsilon_4-(s_5+s_6-t_2)\ \epsilon_5\\
s_1\ \epsilon_1+s_2\ \epsilon_3+(s_2+s_3-t_2)\
\epsilon_4+(s_3+s_6-t_2-t_3)\ \epsilon_5\\
-(s_1+s_2-t_1)\ \epsilon_1-
s_2\ \epsilon_2-s_3\ \epsilon_4-(s_3+s_4-t_3)\ \epsilon_5\\
(s_2+s_5-t_1-t_2)\ \epsilon_1+(s_2+s_3-t_2)\ \epsilon_2+s_3\
\epsilon_3+s_4\ \epsilon_5\\
-(s_5+s_6-t_2)\ \epsilon_1-(s_3+s_6-t_2-t_3)\ \epsilon_2-(s_3+s_4-t_3)\
\epsilon_3-s_4\ \epsilon_4
\end{array}
\ri)=0\ ,
\label{ese}
\ee
where we introduced the vector $v=(v^1,\ v^2,\ v^3,\ v^4,\ v^5)$ with all components vanishing due to $\epsilon\cdot s=0$.
Thus the vanishing of the Gram determinant, $\det s=0$,
ensures self-consistency of the above identity. Although Eq.(\ref{ese}) will be important for
understanding how six-gluon amplitudes transform under cyclic
permutations, it is convenient to keep as many scalars and
pseudoscalars as allowed by momentum conservation, without using the
Gram determinant constraint or Eq.(\ref{ese}) explicitly to eliminate the redundant invariants.

\subsection{\label{int6}Integrals and Their Six-Element Basis}

The integrals \req{FULL} over three vertex positions have the form \req{triple},
with nine integers $n_1,n_2,n_3,n_{11},n_{12},n_{22},n_{13},n_{23},n_{33}$.
Now six functions are necessary to form the integral basis. 
A convenient basis to start with is:
\be
\begin{array}{lcllcllcl}
&F_1=&\ds{F\ts\lf[{3,2,1\atop 0,0,0,0,0,0}\ri]\ ,}
&F_3=&\ds{F\ts\lf[{4,3,2\atop 0,0,0,-1,0,0}\ri]\ ,}
&F_5=&\ds{F\ts\lf[{4,3,2\atop 0,-1,0,-1,0,0}\ri]\ ,}\\[5mm]
&F_2=&\ds{F\ts\lf[{4,3,1\atop 0,-1,0,0,0,0}\ri]\ ,}
&F_4=&\ds{F\ts\lf[{4,4,2\atop 0,-1,0,0,-1,0}\ri]\ ,}
&F_6=&\ds{F\ts\lf[{4,3,2\atop 0,0,0,-1,-1,0}\ri]}\ .
\label{FUNCS}
\end{array}
\ee
In choosing the above functions, we were guided by their low-energy power expansions in  $\ap$, by  their soft limits and by their transformation properties under cyclic permutations.
The $\ap$-expansion of $F_1$, derived in Appendix B, is:
\begin{eqnarray}\label{exR1}
{}F_1&=&\fc{1}{s_2s_6t_2}-\fc{\pi^2}{6}\ \lf(\ \fc{s_4}{s_2s_6}+\fc{s_5}{s_2t_2}+
\fc{s_3}{s_6t_2}\ \ri)\\
&+&\zeta(3)\ \lf(\fc{s_4+s_5-t_1}{s_2}+\fc{s_3+s_4-t_3}{s_6}+\fc{s_4^2+s_4t_2}{s_2s_6}+
\fc{s_5^2+s_5s_6}{s_2t_2}+\fc{s_2s_3+s_3^2}{s_6t_2}\ri)+\ldots\ ,\nonumber
\end{eqnarray}
while the expansion of $F_2$, see also Appendix B, starts  with a single pole:
\be
{}F_2=\fc{\pi^2}{6}\ \fc{1}{s_6}-\zeta(3)\  
\fc{s_2+s_3+s_4+t_2+t_3}{s_6}+\ldots\ .
\label{exR2}
\ee
%Guided by the choice \req{onefinds} for $N=5$ we have chosen a 
%function $F_1$, whose $\ap$--expansion starts with one triple pole in
%the kinematical invariants \req{in} and a 
%second function $F_2$ with a single pole (multiplied by $\fc{\pi^2}{6}$). 
These functions are related by the soft limit $k_6\ra 0$ \cite{appa2} to the
five-gluon functions $f_1$ and $f_2$ of Eq.\req{onefinds}:
\be
s_2 s_6 t_2 F_1\xrightarrow[k_6= 0]{} s_2 s_5 f_1 \ \ ,\ \ \ 
s_6 F_2\xrightarrow[k_6= 0]{} f_2 .
\ee
In fact, the expansions \req{exR1} and \req{exR2} are very similar to \req{rexp} and \req{rexp2}, respectively.
The remaining four functions have no poles.
In particular, the low-energy expansion of $F_3$ begins with the constant $\zeta(3)$:
\be\label{expf3}
F_3=\zeta(3)-\fc{1}{4} \zeta(4) \lf( s_1+4 s_2+3 s_3+2 s_4+3
s_5+4 s_6+t_1+4 t_2+t_3 \ri)+\ldots\ .
\ee
This function is not cyclic invariant; three additional functions, $F_4,F_5$ and $F_6$,
are necessary in order to form a closed representation of $\mathbb{Z}_6$.
Under the generator $i \rightarrow i+1$ mod 6, they transform in the following way:
\be 
F_3\longrightarrow -F_3+F_5\ \ \ ,\ \ \ 
F_6\lra F_5\ \ ,\ \ F_5\lra F_4\ \ ,\ \ F_4\lra F_6\ .
\label{Boston}
\ee
The low-energy expansions of the functions $F_4,F_5$ and $F_6$ also begin with $\zeta(3)$:
\begin{eqnarray}
F_4&=&2 \zeta(3)-\fc{1}{4} \zeta(4) \lf( 7 s_1+5 s_2+5 s_3+7 s_4+5
s_5+5 s_6+5 t_1+2 t_2+5 t_3 \ri)+\ldots\ ,\nonumber \\
F_5&=&2 \zeta(3)-\fc{1}{4} \zeta(4) \lf( 5 s_1+5 s_2+7 s_3+5 s_4+5
s_5+7 s_6+2 t_1+5 t_2+5 t_3 \ri)+\ldots\ ,\hskip1cm \\
F_6&=&2 \zeta(3)-\fc{1}{4} \zeta(4) \lf( 5 s_1+7 s_2+5 s_3+5 s_4+7
s_5+5 s_6+5 t_1+5 t_2+2 t_3 \ri)+\ldots\ . \nonumber
\end{eqnarray}

{}For completeness, we list here also the cyclic transformations of $F_1$ and $F_2$:
\begin{eqnarray}
s_2s_6t_2\ F_1&\ra&s_2s_6t_2\ F_1+s_6\ 
(s_2 s_3 - s_3 s_4 - s_1 s_6 + s_5 s_6 + s_4 t_2 - s_5 t_3)\ F_2\nonumber \\ 
&&- s_3 s_6\ (s_4+s_5-t_1)\ (F_4-F_6)-s_3 t_3\ (s_2+s_5-t_1-t_2)\ (F_3-F_5) \nonumber\\
&&- s_6 t_2\ (s_1+s_4-t_1-t_3)\ F_3-(s_1 s_6-s_5 s_6+s_5 t_3-t_2 t_3)\!\times\label{ftr1} \\ \times&&\hskip -8mm\lf[(s_4+s_5-t_1)(F_4-F_6)+(s_1-s_3+s_5-t_1)(F_3-F_5)
+(s_2+s_4-s_6-t_1)F_3\ri],\nonumber\\
s_6\ F_2&\ra& s_6\ F_2+s_6\ (F_4-F_6)+(s_2-t_1)\
(F_3+F_4-F_5)\nonumber\\  && \qquad\qquad +\ (s_5-t_2)\ (F_3+F_4-F_5-F_6)\label{ftr2}\ .
\end{eqnarray}
Although all six-gluon integrals  can be expressed in terms of the basis \req{FUNCS},
we will see that the actual amplitude involves certain combinations that assume a simpler form when written in the original notation of Eq.\req{triple}. In Appendix C, we will express the relevant integrals in terms of $F_k,~k=1,\dots, 6.$

\subsection{\label{amp6}MHV Amplitude}

The results of  \cite{Oprisa:2005wu} allow expressing the full
six-gluon string amplitude in terms 
of six generalized hypergeometric functions \req{FUNCS}, 
with each function multiplying a 
long combination of kinematic factors involving all possible contractions among the momentum and polarization vectors.
Unlike in the 
five-gluon case, the amplitude cannot be simply separated into parts  
associated to some functions like $T$ and $K_3$ that can be attributed to
distinct effective interactions, see Eq.\req{ampl5}.
Experience with QCD suggests that
the complications are related to the existence of the non-MHV part, with $(---++\,+)$ 
helicity configurations. It is reasonable, however, to expect that the
$(--+++\,+)$ MHV amplitude can be simplified. To that end, we
substitute to the general 
expression \cite{Oprisa:2005wu} the following polarization vectors:
\be  \displaystyle
\xi^-_{\mu}(i)~=~-\frac{\langle
  k_6^+|\gamma_{\mu}|k_i^+\rangle}{\sqrt{2}\,[6\,i\,]} ~
\makebox{for}~i=1,2 \qquad\makebox{and}\qquad
\xi^+_{\mu}(j)~=\quad\frac{\langle
  k_1^-|\gamma_{\mu}|k_j^-\rangle}{\sqrt{2}\,\langle 1j\,\rangle}
 ~\makebox{for}~ j=3,4,5,6.
\label{ximu6}\ee
The kinematic terms surviving in such a configuration contain only one
$\xi_i\xi_j$ factor, and the remaining four polarization vectors are contracted with the
momenta. A generic term has the form
\be
\frac{\langle12\rangle\langle 1j\rangle\langle 1k\rangle\langle 1m\rangle\langle 1n\rangle}{\langle13\rangle\langle 14\rangle\langle 15\rangle\langle 16\rangle}\,\frac{[j6][k6][\alpha 6][\beta m][\gamma n]}{[16][26]},\, \makebox{with}~\,\alpha\neq\beta\neq\gamma=3,4,5;~j,k,m,n=2,3,4,5
\ee
times a linear combination
of the six basis functions, with the coefficients being rational functions of scalar invariants.
There are more than one thousand of such terms, so it is quite a tedious task to simplify
the answer. The final result can be written as
\be
\label{m6}
A(1^-,2^-,3^+,4^+,5^+,6^+)=
\lf[\ V^{(6)}(s_i,t_i)-2i\ \sum_{k=1}^{k=5}\,\epsilon_k\,P^{(6)}_k(s_i,t_i)\ \ri]\ 
\mathfrak{M}_{Y\!M}^{(6)}\ ,\ee
with the functions:
\begin{eqnarray}\nonumber\ts
  P^{(6)}_1&=& s_1\ \ts \FF{4,3,2}{0,-1,0,-1,-1,0}+(s_2+s_5-t_1-t_2)\ \FF{4,3,2}{0,-1,0,-1,0,0}\\[.5mm]\ts
&&\hskip 5mm +~(s_5+s_6-s_1-t_2)\ \ts \FF{4,4,3}{0,0,0,-1,-1,0}\ ,\nonumber \\ \ts
P^{(6)}_2&=& s_2\ \ts\FF{3,3,2}{0,-1,0,0,-1,0}+(s_3+s_6-t_2-t_3)\ \ts\FF{4,4,2}{0,-1,0,0,-1,0}
\ ,\nonumber\\ \ts
P^{(6)}_3&=& s_3\ \ts\FF{4,3,2}{-1,0,0,0,-1,0}+(s_1+s_4-t_1-t_3)\ \ts\FF{4,4,3}{0,0,0,-1,-1,0}\ ,
\label{ps6}\\ \nonumber \ts
P^{(6)}_4&=& s_4\ \ts\FF{4,3,2}{0,0,-1,-1,0,0}+(s_2+s_3-s_4-t_2)\ \ts\FF{4,3,2}{0,0,0,-1,0,0}\ , \\ \nonumber
P^{(6)}_5&=& s_5\ \ts\FF{4,3,2}{0,-1,0,0,0,-1}+(s_3+s_4-s_5-t_3)\ \ts\FF{4,3,2}{0,-1,0,-1,0,0}\ ,\\[5mm]
\nonumber 
\end{eqnarray}
\begin{eqnarray}
V^{(6)}  &=& s_2s_5t_2\ {\ts\FF{3,2,2}{0,0,0,0,0,-1}}+
\frac{1}{2}\ (s_2 s_3 - s_3 s_4 + s_3 s_6 + s_4 t_2 - s_2 t_3 - t_2 t_3)\ P^{(6)}_1\nonumber\\ &&
\hskip -3mm +\,\frac{1}{2}\ 
(-s_2 s_3 + s_1 s_4 - s_4 s_5 - s_3 s_6 + s_3 t_1 - s_4 t_2 
+ s_2 t_3 + s_5 t_3 - t_1 t_3 + t_2t_3)\ P^{(6)}_2\label{vs6}\\ &&\hskip -3mm +\,\frac{1}{2}\ 
(s_2 s_3 - s_1 s_4 + s_2 s_5 + s_3 s_6 + s_5 s_6 - s_3 t_1 - s_6 t_1  - s_2 t_3 - 
  s_5 t_3+ t_1t_2  + t_1t_3 - t_2t_3)\ P^{(6)}_3\nonumber\\ &&\hskip -3mm +\,\frac{1}{2}\ 
(-s_2 s_3 + s_1 s_4 - s_2 s_5 - s_1 s_6 + s_3t_1 + s_6t_1 + s_1t_2+ s_2 t_3 - 
t_1t_2  - t_1t_3)\ P^{(6)}_4\nonumber\\ &&\hskip -3mm +\,\frac{1}{2}\ 
(-s_1 s_2 + s_2 s_3 + s_2 s_5 - s_3 t_1 - s_1 t_2 + t_1t_2)\ P^{(6)}_5-s_5s_3\ P^{(6)}_2+
s_5(s_3-t_2)\ P^{(6)}_3\ . \nonumber
\label{V6}
\end{eqnarray}
The result can be expressed in the basis of functions introduced in Section \ref{int6} by using the formulas written in Appendix C.

Although the six-gluon $V$ and $P$ functions appear complicated, they have very simple 
transformation 
properties under cyclic permutations.
After expressing  them in terms of the basis functions $F_k$, $k=1,\dots, 6$, see Appendix C, and using the transformation properties (\ref{Boston}, \ref{ftr1}, \ref{ftr2}), it is easy to see that $V^{(6)}(s_i,t_i)$ is cyclic invariant. Furthermore,
the functions $P^{(6)}(s_i,t_i)$ transform among themselves 
in such a way that the imaginary part
of the Yang-Mills formfactor in Eq.\req{m6} is also invariant. This can be seen in the following way. Let us put
the five functions $P^{(6)}_i$ into the vector
$P^{(6)}=(P^{(6)}_1,\ -P^{(6)}_2,\ P^{(6)}_3,\ -P^{(6)}_4,\ P^{(6)}_5)$.
Then the action of $i\to i+1$ mod 6 on $\epsilon$ and 
$P^{(6)}$ can be written as
\be
\epsilon\longrightarrow \epsilon\ M\ \ \ ,\ \ \ 
P^{(6)}\longrightarrow P^{(6)}\ (M^{t})^{-1}+\Delta\ F_3\ ,
\ee
with the unimodular matrix $M$ and the vector $\Delta$:
\be
M=\lf(\begin{matrix}
1&1&1&1&1\\
-1&0&0&0&0\\
0&-1&0&0&0\\
0&0&-1&0&0\\
0&0&0&-1&0
\end{matrix}\ri)\ \ \ ,\ \ \ 
\Delta^t=\lf(\begin{matrix}
-s_1 - s_2 + s_6 + t_1\\
-s_1 + s_2 - s_6 + t_3\\
s_1 + s_4 - t_1 - t_3\\
s_3 - s_4 - s_5 + t_1\\
-s_3 - s_4 + s_5 + t_3
\end{matrix}\ri)\ .
\ee
However, with Eq.(\ref{ese}), we find: 
\be 
\epsilon\ \Delta^t=v^1+v^2+v^4+v^5=0\ ,
\ee
thus $P^{(6)}\ \epsilon^t\rightarrow P^{(6)}\ \epsilon^t$.
As a result, similarly to the case
of four and five gluons, the full string formfactor of the MHV six-gluon 
amplitude (\ref{m6}) is cyclic invariant. 
It also has the correct soft limits \cite{appa2} when any momentum goes to zero:
\begin{eqnarray}\label{factorize}
V^{(6)}(s_i,t_i)&\xrightarrow[k_j= 0]{}&V^{(5)}(s_j)\\
\sum_{l=1}^5 (-1)^{l+1}\ P^{(6)}_l(s_i,t_i)\xrightarrow[k_6= 0]{}P^{(5)}(s_j)\, , && \!\!\!
P^{(6)}_l(s_i,t_i)\xrightarrow[k_l= 0]{}P^{(5)}(s_j)\quad\makebox{for}\quad l\leq 5\ . \nonumber
\end{eqnarray}
%factorizing into the infrared pole times the five-gluon amplitude
%(\ref{m5}). 
{}Furthermore, it has the right
collinear limits \cite{appa2}, when the momenta of adjoining gluons, $k_i$ and $k_{i+1}$, with $i+1$ 
mod 6, become parallel:
\begin{eqnarray}\label{COLL}
V^{(6)}(s_i,t_i)&\xrightarrow[k_i||k_{i+1}]{}&                V^{(5)}(s_j)\ ,\nonumber \\
\sum_{k=1}^5\epsilon_k\,P^{(6)}_k(s_i,t_i)&\xrightarrow[k_i||k_{i+1}]{}&  
\epsilon(1,2,3,4)\ P^{(5)}(s_j)\ .
\end{eqnarray}

The low-energy 
behavior of the amplitude  (\ref{m6}) is determined, up to the order
${\cal O}(\alpha'^3)$, by the following expansions:
\begin{eqnarray}\label{lowlimits}\nonumber
V^{(6)}(s_i,t_i)&\approx& 1-\frac{\pi^2}{12}\ (\ \{s_1s_2\}-\{s_1s_4\}+\{t_1t_2\}\ )+
\frac{\zeta(3)}{2}\ \left( \{s_1s_2^2\}+\{s_1^2s_2\}-\{s_1^2s_4\}\right.\\ \nonumber
&&\hskip -2cm\left. +\{s_1s_2t_1\}
-\{s_1s_4t_1\}-\{s_2s_5t_1\}-3 \{s_1s_4t_2\}+\{s_1t_1t_3\}+\{t_1t_2^2\}+\{t_1^2t_2\}+3\, t_1t_2t_3
\right),\\ \nonumber
P^{(6)}_1(s_i,t_i)&\approx&
\frac{\pi^2}{6}+\zeta(3)\ (s_1 + 2 s_2 - s_3 - s_4 + 2 s_5 + s_6 - 3 t_1 - 3 t_2 - t_3)
\ ,\\ 
 P^{(6)}_2(s_i,t_i)&\approx&
\frac{\pi^2}{6}+\zeta(3)\  (2 s_2 + 2 s_3 - s_4 - s_5 + s_6 - t_1 - 3 t_2 - 2 t_3)
\ ,\\ \nonumber
P^{(6)}_3(s_i,t_i)&\approx&
\frac{\pi^2}{6}+\zeta(3)\ (2 s_3 + s_4 - s_5 - s_6 - t_1 - t_2 - 2 t_3)
\ ,\ \\ \nonumber
P^{(6)}_4(s_i,t_i)&\approx&
\frac{\pi^2}{6}+\zeta(3)\ (-s_1 + s_3 + s_4 - s_6 - t_1 - t_2 - t_3)\ ,
\\ \nonumber
P^{(6)}_5(s_i,t_i)&\approx&
\frac{\pi^2}{6}+\zeta(3)\ (-s_1 - s_2 + s_3 + 2 s_4 - t_1 - t_2 - 2 t_3)\ .
\end{eqnarray}

All other partial amplitudes $A(\sigma(1^-),\sigma(2^-),\sigma(3^+),\sigma(4^+),\sigma(5^+),\sigma(6^+))$ can be obtained from $A(1^-,2^-,3^+,4^+,5^+,6^+)$ by simply applying the 
$S_6/\mathbb{Z}_6 $ coset permutations $\sigma$ to the right hand side of Eq.\req{m6}.

 \subsection{\label{recon}Reconstructing the Amplitude from First Principles}
After checking that the amplitude \req{m6} satisfies all self-consistency conditions
following from cyclic symmetry and soft/collinear limits,
we would like to proceed in reverse, in order to understand to what extent the form of the string factor is determined by these  conditions. To that end, we make 
the following ansatz for the function  $V^{(6)}(s_i,t_i)$:
%Stephan, this is in English! -- nouns are lower case
\be
\widetilde V^{(6)}(s_i,t_i)=s_2\ s_6\ t_2\ F_1+   (s_6\ F_2)\ \sum_{1\leq i\leq j\leq 9}\lambda_{ij} \ 
s_i\ s_j+\sum_{l=3}^6 F_l\sum_{1\leq i\leq j\leq k\leq 9}\lambda^l_{ijk}\ s_i\ s_j\ s_k\ ,
\label{ansatz1}
\ee
in the basis of six functions \req{FUNCS}, with $45$ and $660$ real constant coefficients $\lambda_{ij}$ 
and $\lambda_{ijk}^l$, respectively. We will try to fix these constants
by imposing the above self-consistency conditions. For convenience, we 
use the notation $s_7\equiv t_1,\ s_8\equiv t_2,\ s_9\equiv t_3$.
The leading term in the $\ap$ expansion of the ansatz \req{ansatz1} which, according to Eq.\req{exR1}, is equal to 1, is dictated by the zero slope (Yang-Mills) limit,
$V^{(6)}(s_i,t_i)\approx 1$.
The next-to-leading order ${\cal O}(\ap^2)$, with the common $\zeta(2)=\fc{\pi^2}{6}$ factor, is governed by the constants $\lambda_{ij}$. 
They are completely determined by the cyclic symmetry and soft limits.
Finally, after imposing the right collinear limits, 
all remaining constants $\lambda_{ijk}^l$ can be expressed in terms of one of them, 
$\lambda_{789}^4$.
In this way, the ansatz \req{ansatz1} becomes
\be
\widetilde V^{(6)}(s_i,t_i)=V^{(6)}(s_i,t_i)+
\lambda_{789}^4\ \lf(\ t_1t_2t_3-\{\ s_1s_4t_2\ \}\ \ri)\ (F_4+F_5+F_6)\ ,\ee
thus the real part of the string factor is completely determined by the self-consistency conditions, up to one constant. The respective term is cyclic invariant and vanishes in both soft and collinear limits.

In order to examine the imaginary part of the string factor, we assume that it has the form
$\sum\limits_{m=1}^5\epsilon_m \widetilde P^{(6)}_m(s_i,t_j)$,
with the following ansatz
\be
\widetilde P^{(6)}_m(s_i,t_i)=s_6\ F_2+\sum_{l=3}^6 F_l\ \sum_{k=1}^9 \mu^l_{mk}\ s_k
\label{ansatz2}
\ee
for the five functions $P^{(6)}_m(s_i,t_i)$.
Here again, we try to fix 180 real constants $\mu^l_{mk}$ by demanding that
the sum $\sum\limits_{m=1}^5\epsilon_m \ \widetilde P^{(6)}_m$ be cyclic invariant and that it has the correct soft/collinear limits. The latter requirement fixes $156$ of  $180$ 
constants. By further imposing cyclic invariance, one ends up with only four arbitrary constants, $\mu_{56}^4,\ \mu_{57}^4,\ \mu_{58}^4$ and $\mu_{58}^3$. Finally,
after using the relations between pseudoscalar invariants, written as $v=0$ in Eq.\req{ese}, one finds
\comment{Furthermore after imposing cyclic invariance, additional $20$ coefficients $\mu^l_{mk}$
are fixed. Eventually the Ansatz \req{ansatz2} becomes
the imaginary part of the bracket $[]$ in (\ref{m6})
\be
\begin{array}{lcl}
\ds{\sum_{m=1}^5\epsilon_m \tilde P^{(6)}_m(s_i,t_i)}&=&\ds{
\sum_{m=1}^5\epsilon_m P^{(6)}_m(s_i,t_i)
+\fc{1}{6}\ (v^1+v^2+v^4+v^5)\  (F_6-2F_5-F_4-2F_3)} \\
&&\hskip-2.5cm\ds{\fc{1}{6}\ (v^2+v^5)\ (F_6+F_5+2F_4-2F_3)+\lf(\mu_{58}^3+\fc{1}{3}\ri)\ 
(v^1-4v^2-3v^3-2v^4-v^5)\ F_3} \\
&&\hskip-1.5cm\ds{+\lf(\mu_{58}^3+\fc{1}{3}\ri)\ \lf[(2 v^2+v^3)\ 
F_6+(-v^1+v^2+v^3+v^4+v^5)\ F_5+(v^1-v^2)\ F_4\ri]} \\
&-&\ds{(v^1 +v^2 +v^3)\ 
\lf[\lf(\mu_{56}^4-\fc{5}{6}\ri)\ F_6-\lf(\mu_{57}^4+\fc{5}{6}\ri)\ F_4-
\lf(\mu_{58}^4+\fc{2}{3}\ri)\ F_5)\ri]} \\
&-&\ds{(v^3 + v^4+v^5)\ 
\lf[\lf(\mu_{56}^4-\fc{5}{6}\ri)\ F_4-\lf(\mu_{57}^4+\fc{5}{6}\ri)\ F_5-
\lf(\mu_{58}^4+\fc{2}{3}\ri)\ F_6) \ri]} \\
&+&\ds{(v^2+v^3+v^4)\ 
\lf[\lf(\mu_{56}^4-\fc{5}{6}\ri)\ F_5-\lf(\mu_{57}^4+\fc{5}{6}\ri)\ F_6-
\lf(\mu_{58}^4+\fc{2}{3}\ri)\ F_4)\ri]} \\
&&\hskip-1cm\ds{-3\ \lf(\mu_{58}^3+\fc{1}{3}\ri)\ \lf[(s_6 - t_2)\ \epsilon_5 + (s_2 - t_2)\ 
\epsilon_4 + s_2\ \epsilon_3\ri]\ (F_6+F_5-F_4-2\ F_3)\ ,}
\end{array}\label{map2}
\ee
up to some extra terms involving the remaining four unconstrained parameter 
$\mu_{56}^4, \mu_{57}^4,\ \mu_{58}^4$ and $\mu_{58}^3$.
According to \req{ese} all terms comprising components $v^i$ of the vector $v$ vanish.
Hence in \req{map2} only the last term (proportional to $\mu_{58}^3$) is yet unfixed:}
\begin{eqnarray}
\sum_{m=1}^5\epsilon_m \widetilde P^{(6)}_m(s_i,t_i)&=&\sum_{m=1}^5\epsilon_m P^{(6)}_m(s_i,t_i)
\nonumber\\
&-&\lf(3\ \mu_{58}^3+1\ri)\ \lf[(s_6 - t_2)\ \epsilon_5 + (s_2 - t_2)\ 
\epsilon_4 + s_2\ \epsilon_3\ri]\ (F_6+F_5-F_4-2\ F_3)\nonumber\ ,
\end{eqnarray}
thus also the imaginary part of the string factor is determined up to one constant.

There is one more constraint available. In the Abelian case, the leading term in the $\ap$ expansion must vanish because it is entirely due to Yang-Mills gluon self-interactions.
The next-to-leading term, which is associated to the ${\cal I}_{F^4}$ interaction, must also vanish: while four Abelian gauge bosons interact via the corresponding Born-Infeld term, they cannot spread via Yang-Mills interactions, like in the left diagram on Figure 1. By requiring that the $\ap$-expansion of the Abelian amplitude
starts at order higher than ${\cal O}(\ap^2)$,
we obtain $\lambda_{789}^4=0$ and $\mu_{58}^3=-1/3$ \cite{prep}. Actually, with this constraint, 
the Abelian amplitude starts at order ${\cal O}(\ap^4)$ [with the common factor of $\zeta (4)$].
We conclude that the six-gluon MHV amplitude can be uniquely determined from first principles.
It is worth mentioning that in the simpler five-gluon case, the cyclic symmetry and soft limit are completely sufficient to determine the amplitude, similarly to the constants $\lambda_{ij}$ that govern the next-to-leading ${\cal O}(\ap^2)$ contribution
to the six-gluon amplitude. The reason, to be elaborated in the next section, is that all
${\cal O}(\ap^2)$ terms originate from ${\cal I}_{F^4}$ interactions.

\section{\label{ngluons}$\bm{N}$ Gluons}
It is clear from the discussion of $N=5$, and especially of $N=6$, that the computational complexity  increases steeply with $N$. The integrals \req{FULL} become more complicated and the number of independent functions grows. The functions emerging in the step from $N{-}1$ to $N$ have low-energy expansions starting at ${\cal O}(\alpha'^{N-3})$ [with a common factor of $\zeta(N{-}3)$]. On the other hand, the simple, factorized forms of Eqs.\req{m4}, \req{m5} and \req{m6} strongly suggest that MHV configurations enjoy a special status. The fact that $N=5$ as well as $N=6$ MHV amplitudes can be reconstructed 
from first principles, by using very simple physical constraints, is very encouraging because it opens way
to an iterative procedure suitable for larger numbers of gluons. It also indicates the existence of some recursion relations similar to those in QCD \cite{Berends:1987me}. A recursive construction of the amplitudes requires however a better
understanding of the space of generalized hypergeometric integrals
\req{FULL}. This ingredient will have to wait until completion of
Ref. \cite{prep}. Nevertheless, already at this point, we can determine
the leading ${\cal O}(\alpha'^2)$ string corrections.

An $N$-gluon scattering process can be parameterized in terms of $N(N-3)/2$
kinematic invariants which can be chosen as the cyclic orbits of $\textstyle
\bl 1\br_k,~k=1,\dots,E(\frac{N}{2}-1)$, where $E$ denotes the integer part.
Recall that the cyclic $\mathbb{Z}_N$ group is generated by the shift of indices labeling gluons from $i\to i+1$ mod $N$.
Note that for $N$ odd, the last orbit contains $N$ elements, while for $N$ even their number
is reduced  by the momentum conservation to $N/2$. As in the case of $N=6$, we can ignore the four-dimensional Gram determinant constraints \cite{asribekov} that reduce the number of independent invariants to $3N-10$. We also keep the pseudoscalars
$\epsilon(k,l,m,n)$, with $k<l<m<n<N$, which are independent as far as the momentum
conservation is concerned but are related by equations similar to (\ref{ese}).

Let us first collect all known ${\cal O}(\alpha'^2)$ terms and rewrite the leading and next-to-leading terms in the low-energy expansions of
$N=4,5,6$ MHV amplitudes as
\be\label{mn}
A(1^-,2^-,3^+,4^+,\cdots,N^+)=\lf[\ 1-\frac{\pi^2}{12}\ Q^{(N)}\ \ri]\
\mathfrak{M}_{Y\!M}^{(N)}\ +\ {\cal O}(\ap^3)\ ,\ee
where $Q^{(N)}$ are the following  Lorentz-invariant, homogenous of degree four, functions of the momenta:
\begin{eqnarray}Q^{(4)}&=& s_1s_2\nonumber\\ Q^{(5)}&=&s_1s_2+s_2s_3+s_3s_4+s_4s_5+s_5s_1+4i\,\epsilon(1,2,3,4) \\ Q^{(6)}&=&s_1s_2+s_2s_3+s_3s_4+s_4s_5+s_5s_6+s_6s_1+t_1t_2+t_2t_3
+t_3t_1-s_1s_4-s_2s_5-s_3s_6
\nonumber\\
&&+~4i\,[\epsilon(1,2,3,4)
+\epsilon(1,2,3,5)+\epsilon(1,2,4,5)+\epsilon(1,3,4,5)+\epsilon(2,3,4,5)]
\nonumber\end{eqnarray}
\begin{figure}
\includegraphics[scale=0.7]{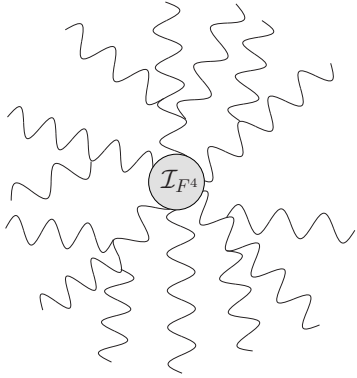}
\caption{\label{fig:mhvfig2} Feynman diagrams contributing the leading ${\cal O}(\alpha'^2)$ string corrections to $N$-gluon YM amplitudes involve one ${\cal I}_{F^4}$ effective interaction vertex, while the remaining vertices are due to the tree-level YM interactions.}
\end{figure}
At this order, the corrections are generated by gluonic tree diagrams involving only one ${\cal I}_{F^4}$  vertex, \cf Eq.\req{if4}, and a number of standard Yang-Mills interactions.
A typical diagram contributing to $N$-gluon scattering is shown in Figure 2.
In order to determine $Q^{(N)}$ for arbitrary $N$ one can either calculate the sum of such Feynman diagrams or apply iteration utilizing the soft limit and $\mathbb{Z}_N$ symmetry.
Although it is possible to formulate the latter as a formal recursion relation, we
prefer to apply the iterative procedure explicitly, step by step, starting from $Q^{(7)}$.
Here again, it will be very convenient to use the notation introduced in Section \ref{five}: an expression enclosed
inside curly brackets $\{\cdots\}$ denotes the sum over all distinct elements
of its $\mathbb{Z}_N$ cyclic permutation group orbit. Thus, for example, 
\be Q^{(6)}=\{s_1s_2\}+\{t_1t_2\}-\{s_1s_4\}+~4i\hskip -4mm\sum_{k<l<m<n<6}\!\!\!\epsilon(k,l,m,n)\, .\ee
 
{}For $N=7$, the 14 invariants are $s_i\equiv\bl i\br_1,~i=1,\dots,7$ and $t_j\equiv\bl j\br_2,~j=1,\dots,7$. The real part of $Q^{(7)}$ must
be a cyclic invariant, quadratic form in $s_i,\,t_j$. There are 15 quadratic cyclic invariants,
but there exists only one linear combination,
\be\makebox{Re} Q^{(7)}=\{s_1s_2\}+\{t_1t_2\}-\{s_1t_4\}\ee
that gives $\makebox{Re} Q^{(6)}$ in the soft  limit
\be k_7\to 0:\qquad s_6\to 0,\quad s_7\to 0,\quad  t_4\to t_1,\quad  t_5\to s_5,\quad  t_6\to s_6,\quad  t_7\to s_1.\ee
In order to determine the imaginary part, we first note that although there are 15 linearly independent pseudoscalars, there are only 3 $\mathbb{Z}_7$-invariant combinations. We list them below, together with their $k_7\to 0$ limits:
\begin{eqnarray}\nonumber\epsilon(1,2,3,4)\!
&+&\!\epsilon(1,2,3,6)+\epsilon(1,2,5,6)+\epsilon(1,4,5,6)+\epsilon(3,4,5,6)
\longrightarrow\epsilon(2,3,4,5),\\ \nonumber
\epsilon(1,2,3,5)\!
&+&\!\epsilon(1,2,4,5)+\epsilon(1,3,5,6)+\epsilon(2,3,4,5)+\epsilon(2,4,5,6)
\longrightarrow\epsilon(1,3,4,5)+\epsilon(2,3,4,5),\\
\nonumber\epsilon(1,2,4,5)\!
&+&\!\epsilon(1,3,4,5)+\epsilon(1,3,4,6)+\epsilon(2,3,4,6)+\epsilon(2,3,5,6)
\longrightarrow\epsilon(1,2,3,5)+\epsilon(1,2,4,5).
\end{eqnarray}
Thus the imaginary part of $Q^{(7)}$ is also uniquely determined to be:
\be\makebox{Im} Q^{(7)}=\,4\!\!\!\!\!\sum_{k<l<m<n<7}\!\!\!\epsilon(k,l,m,n)\, .\ee
The above iteration can be continued to a larger number of gluons,  with the unique answer:
\begin{eqnarray}\nonumber
Q^{(N)}&=& \sum_{k=1}^{E(\frac{N}{2}-1)}\{\,\bl 1\br_k\bl 2\br_k\}\,-\!\!
\sum_{k=3}^{E(\frac{N}{2}-1)}\{\,\bl 1\br_k\bl 2\br_{k-2}\}\,+\,C^{(N)}+\, 4i \hskip -4mm \sum_{k<l<m<n<N}\!\!\!\epsilon(k,l,m,n)\, ,\\ && \label{mhvn} \\[-2mm]
&&\hskip -6mm C^{(N)} = \left\{ \begin{array}{ll}
        -\{\,\bl 1\br_{\frac{N}{2}-2} \bl \frac{N}{2}+1\br_{\frac{N}{2}-2}\}  & 
\mbox{$N>4,$ even}\\[2mm]
         -\{\,\bl 1\br_{\frac{N-5}{2}} \bl \frac{N+1}{2}\br_{\frac{N-3}{2}}\} & 
\mbox{$N>5,$ odd}.\end{array}\nonumber \right.
\end{eqnarray}

It is very interesting that Eq.(\ref{mhvn}) bears a striking resemblance to the one-loop all positive helicity amplitudes of QCD \cite{Bern:1993qk}.\comment{If one formally replaces 
$\langle 12\rangle^4$ by $1/\alpha'^{2}$ in $\mathfrak{M}_{Y\!M}^{(N)}$, the leading string correction to this ``mostly plus'' MHV amplitude becomes identical to Eqs.(6)-(8) of Ref.\cite{Bern:1993qk}, up to an overall momentum-independent factor.}
The resemblance originates at the five-gluon level, \cf $\pi^2$ terms in our Eq.\req{v5ex} {\it vis-\`a-vis} Eq.(5) of Ref.\cite{Bern:1993qk}, and then propagates to $N$ gluons because in both cases the multi-gluon results are uniquely determined by the permutation symmetry and soft limits.

Work on the recursive construction of MHV amplitudes to all orders in $\ap$ is in progress 
\cite{prep}.

\section{\label{conclusions}Conclusions}

The main result of this paper, in addition to the specific formulas for scattering amplitudes, 
is the demonstration of a striking simplicity hidden in multi-gluon scattering, persisting 
at the full-fledged string level. The maximally helicity violating  configurations retain 
their special status even after the tree diagrams are replaced by a disk world-sheet.
The string effects are succinctly summarized in a number of kinematic functions, extending 
the well-known result for four gluons to an arbitrary number of gluons. We argued that the 
soft and collinear factorization properties, combined with the Abelian limit, are completely sufficient to determine all $N$-gluon MHV
amplitudes,
however a completely recursive construction requires a better understanding of the boundary integrals determining the kinematic functions. 
Work in  this direction is in progress \cite{prep}.

In superstring theory, it is often possible to describe a single physical process in several 
ways, by using various dualities. We believe that the simplicity 
of MHV amplitudes reflects the existence of an underlying integrable structure not only in QCD, 
but also in type I string theory. Hence it would be very interesting to understand if there 
is any room in the twistor formulation of string theory \cite{Witten:2003nn} that would allow 
accommodating
open string corrections to YM scattering amplitudes. 

Another duality relevant to the  present work is the type I--heterotic duality 
\cite{Witten:1995ex,Polchinski:1995df}. Here, the disk-level interactions of $2n$ gauge field 
strength tensors in type I theory are dual to  the heterotic $(n-1)$-loop interactions. 
In this context, it would be interesting to investigate a possible relation of our results to Ref.\cite{Bern:1993qk} and to the recent computations of all one-loop MHV amplitudes in QCD \cite{Berger:2006vq}. 
Our results should also help in explaining why the heterotic six-gluon amplitudes  are not compatible at the two-loop level with a semi-classical type I description in terms of a  na\"{i}ve extension of Born-Infeld electrodynamics \cite{Stieberger:2002fh,Stieberger:2002wk}.
\comment{In 2002 \cite{Stieberger:2002fh,Stieberger:2002wk}, 
we explicitly computed a subclass of heterotic six-gluon amplitudes at the two-loop level and 
discovered an intriguing discrepancy with type I theory, already in the Abelian subsector of 
the gauge group, where type I interactions are governed by the Born-Infeld action.} With some more
 work, the six-gluon amplitudes presented here could be used to extract the complete type I effective action and could shed more light on the long-standing 
problem how to construct a non--Abelian  generalization of the Born-Infeld Lagrangian.

When looking at the simple formulas describing multi-gluon superstring scattering, incorporating 
infinitely many interactions among infinite number of particles, one cannot stop wondering if 
the effective field theory is really the right framework for describing low-energy string physics. 
For instance, the five-gluon amplitude (\ref{m5}) was used in \cite{Barreiro:2005hv} to 
determine the complete ${\cal O}(\alpha'^3)$ string corrections. The effective action consists 
of hundreds of terms and does not give justice to Eq.(\ref{m5}). The advantage of using the 
effective field theoretical description is that, in principle, it allows going off-shell and 
studying the modifications of classical field equations. However, it is certainly not the most 
efficient way of recording the S-matrix. Historically, string theory grew out from S-matrix 
theory but its formalism has evolved  more and more towards Lagrangian quantum field theory. 
We need a better formalism, somewhere halfway between S-matrix and Lagrangians.

\begin{acknowledgments}
%\vskip-0.5cm
We are grateful to Lance Dixon for very useful correspondence.
In addition, we wish to thank K.S.\ Narain and Dan Oprisa for discussions.
This work  is supported in part by the European Commission under Project MRTN-CT-2004-005104.
The research of T.R.T.\ is supported in part by the U.S.
National Science Foundation Grants PHY-0242834 and PHY-0600304.
T.R.T.\ is grateful to Dieter L\"{u}st,
to Arnold Sommerfeld Center for Theoretical Physics at Ludwig Maximilians
University, and to Max-Planck-Institut in Munich, for their kind hospitality. 
He would like to also thank the Galileo Galilei Institute for Theoretical Physics for 
hospitality and INFN for partial support during completion of this work.
The Feynman diagrams have been created by using the program JaxoDraw \cite{Jaxo}.
Any opinions, findings, and conclusions or
recommendations expressed in this material are those of the authors and do not necessarily
reflect the views of the National Science Foundation.
\end{acknowledgments}

\appendix

\section{\label{appendixA}Kinematic Invariants, Soft Limits and\\ \hspace*{3.6cm} Collinear Limits}
\subsection{Kinematic Invariants}\noindent
The tables below contain the scalar products 
$s_{ij}\equiv 2\ap\ k_ik_j$, with $i$ and $j$ labeling rows and columns, expressed in terms of the kinematic invariants of type \req{in},
used in the paper to describe multi-gluon scattering processes, for $N=4, \ 5$ and 6 gluons.
\begin{center}$\bm{N=4}$\\ \begin{tabular}{p{1cm}p{1cm}p{1cm}p{1cm}p{1cm}}
 \cline{2-4}
  \multicolumn{1}{c}{} &\multicolumn{1}{|c}{2} &\multicolumn{1}{|c}{3}& \multicolumn{1}{|c}{4}&\multicolumn{1}{|c}{}\\ \hline
 \multicolumn{1}{|c}{1} &  \multicolumn{1}{|c}{ $\,\;\quad s_1\quad\;\,$ } &  \multicolumn{1}{|c}{$-s_1-s_2$} &  \multicolumn{1}{|c}{$s_2$} &  \multicolumn{1}{|c|}{1} \\  \cline{1-5}
\multicolumn{1}{|c}{2}& \multicolumn{1}{|c}{} &  \multicolumn{1}{|c}{$s_2$} &  \multicolumn{1}{|c}{$-s_1-s_2$} &  \multicolumn{1}{|c|}{2} \\ 
\hline
 \multicolumn{1}{|c}{3}&\multicolumn{2}{|c}{} &  \multicolumn{1}{|c}{$s_1$} &  \multicolumn{1}{|c|}{3}  \\ 
 \hline
\end{tabular}\end{center}   

\begin{center}$\bm{N=5}$\\ \begin{tabular}{p{1cm}p{1cm}p{1cm}p{1cm}p{1cm}p{1cm}}
 \cline{2-5}
  \multicolumn{1}{c}{} &\multicolumn{1}{|c}{$\,\;\;\qquad 2\qquad\;\;\,$} &\multicolumn{1}{|c}{3}& \multicolumn{1}{|c}{4}& \multicolumn{1}{|c}{5}&\multicolumn{1}{|c}{}\\ \hline
 \multicolumn{1}{|c}{1} &  
\multicolumn{1}{|c}{$s_1$}&\multicolumn{1}{|c}{$ -s_1 - s_2 + s_4$}&  \multicolumn{1}{|c}{$ s_2 - s_4 - s_5$}& \multicolumn{1}{|c}{$s_5$}
&  \multicolumn{1}{|c|}{1} \\  \hline
\multicolumn{1}{|c}{2} &  \multicolumn{1}{|c}{} &  
 \multicolumn{1}{|c}{$s_2$}&\multicolumn{1}{|c}{$ -s_2 - s_3 + s_5$}& \multicolumn{1}{|c}{$-s_1 + s_3 - s_5$}
& \multicolumn{1}{|c|}{2} \\ 
\hline
 \multicolumn{1}{|c}{3} &\multicolumn{2}{|c}{} &  
 \multicolumn{1}{|c}{$ s_3$}&\multicolumn{1}{|c}{$ s_1 - s_3 - s_4$}
&  \multicolumn{1}{|c|}{3}  \\ 
 \hline
\multicolumn{1}{|c}{4} & \multicolumn{3}{|c}{}&  \multicolumn{1}{|c}{$s_4$} &  \multicolumn{1}{|c|}{4} \\ \hline
\end{tabular}\end{center} 
\begin{center}$\bm{N=6}$\\\begin{tabular}{p{1cm}p{1cm}p{1cm}p{1cm}p{1cm}p{1cm}p{1cm}}
 \cline{2-6}
  \multicolumn{1}{c}{} &\multicolumn{1}{|c}{$\;\;\;\qquad 2\qquad\;\;\;$} &\multicolumn{1}{|c}{3}& \multicolumn{1}{|c}{4}& \multicolumn{1}{|c}{5}&\multicolumn{1}{|c}{6}&\multicolumn{1}{|c}{}\\ \hline
 \multicolumn{1}{|c}{1} &  
\multicolumn{1}{|c}{$s_1$}& \multicolumn{1}{|c}{$\;\; -s_1 - s_2 + t_1\;\;$}&\multicolumn{1}{|c}{$ s_2 + s_5 - t_1 - t_2$}&\multicolumn{1}{|c}{$ -s_5 - s_6 + t_2$}&\multicolumn{1}{|c}{$ s_6$}
&  \multicolumn{1}{|c|}{1} \\  \hline
 \multicolumn{1}{|c}{2} &\multicolumn{1}{|c}{} &  
 \multicolumn{1}{|c}{$s_2$}&\multicolumn{1}{|c}{$ -s_2 - s_3 + t_2$}&\multicolumn{1}{|c}{$ s_3 + s_6 - t_2 - t_3$}& \multicolumn{1}{|c}{$-s_1 - s_6 + t_3$}
& \multicolumn{1}{|c|}{2} \\ 
\hline  \multicolumn{1}{|c}{3} &
 \multicolumn{2}{|c}{} & 
\multicolumn{1}{|c}{$ s_3$}& \multicolumn{1}{|c}{$  -s_3 - s_4 + t_3$}& \multicolumn{1}{|c}{$  s_1 + s_4 - t_1 - t_3$} 
&  \multicolumn{1}{|c|}{3}  \\ 
 \hline  \multicolumn{1}{|c}{4}&
 \multicolumn{3}{|c}{}&
  \multicolumn{1}{|c}{$ s_4$}&  \multicolumn{1}{|c}{$-s_4 - s_5 + t_1$}
 &  \multicolumn{1}{|c|}{4} \\ \hline  \multicolumn{1}{|c}{5}&
\multicolumn{4}{|c}{}& \multicolumn{1}{|c}{$ s_5$}&  \multicolumn{1}{|c|}{5} \\ \hline
\end{tabular}\end{center}

\subsection{Soft Limits}\noindent
\def\eps{\epsilon}
The soft limit is defined as $k_i\to 0$ for some $i$. {}For a cyclic invariant function of the momenta, it is sufficient to consider only one soft momentum, say $k_N\to 0$. Then the scalar
invariants describing $N$-gluon kinematics have the following limits in terms of the invariants describing $N-1$ gluons, for $N=5$ and 6:
\begin{center}$\bm{N=5}$\\
\begin{tabular}{|c|c|c|c|c|c|}
\hline
 &$\ s_1\ $ & $\ s_2\ $ & $\ s_3\ $ & $\ s_4\ $ & $\ s_5\ $ \\
\hline
\comment{$\ k_1\ra 0\ $&$0$& $s$ & $u$ & $s$ & $0$\\
$\ k_2\ra 0\ $&$0$ & $0$ & $u$ & $s$ & $u$\\
$\ k_3\ra 0\ $&$s$ & $0$ & $0$ & $s$ & $u$\\
$\ k_4\ra 0\ $&$s$ & $u$ & $0$ & $0$ & $u$ \\ }
$\ k_5\ra 0\ $&$s_1$ & $s_2$ & $s_1$ & $0$ & $0$\\
\hline
\end{tabular}
\end{center}
\begin{center}$\bm{N=6}$\\
\begin{tabular}{|c|c|c|c|c|c|c|c|c|c|}
\hline
 & $\ s_1\ $ & $\ s_2\ $ & $\ s_3\ $ & $\ s_4\ $ & $\ s_5\ $ & $\ s_6\ $ & 
$\ t_1\ $ & $\ t_2\ $ & $\ t_3\ $\\
\hline
\comment{$\ k_1\ra0\ $&$0$ & $s_1$ & $s_2$ & $s_3$ & $s_4$ & $0$ & $s_1$ & $s_4 $ & $s_5$\\
$\ k_2\ra0\ $&$0$ & $0$ & $s_2$ & $s_3$ & $s_4$ & $s_5$ & $s_1$ & $s_2$ & $s_5$\\
$\ k_3\ra0\ $&$s_1$ & $0$ & $0$ & $s_3$ & $s_4$ & $s_5$ & $s_1$ & $s_2$ & $s_3$ \\
$\ k_4\ra0\ $&$ s_1$ & $ s_2$ & $0$ & $0$ & $s_4$ & $s_5$ & $s_4$ & $s_2$ & $s_3$\\
$\ k_5\ra0\ $&$ s_1$ & $ s_2$ & $s_3$ & $0$ & $0$ & $s_5$ & $s_4$ & $s_5$ & $s_3$\\}
$\ k_6\ra0\ $&$ s_1$ & $ s_2$ & $s_3$ & $s_4$ & $0$ & $0$ & $s_4$ & $s_5$ & $s_1$\\
\hline
\end{tabular}
\end{center}
As $k_5\to 0$, the five-gluon Levi-Civita pseudoscalar invariant $\eps(1,2,3,4)\equiv\varepsilon\to 0$. {}For $N=6$, the soft limits of pseudoscalar invariants defined in Eq.\req{pseudo6} are written below:
\begin{center}$\bm{N=6}$\\
\begin{tabular}{|c|c|c|c|c|c|}
\hline
 &$\ \eps_1\ $ & $\ \eps_2\ $ & $\ \eps_3\ $ & $\ \eps_4\ $ & $\ \eps_5\ $ \\
\hline
\comment{$\ k_1\ra 0\ $&$\eps$& $0$ & $0$ & $0$ & $0$\\
$\ k_2\ra 0\ $&$0$ & $-\eps$ & $0$ & $0$ & $0$\\
$\ k_3\ra 0\ $&$0$ & $0$ & $\eps$ & $0$ & $0$\\
$\ k_4\ra 0\ $&$0$ & $0$ & $0$ & $-\eps$ & $0$ \\
$\ k_5\ra 0\ $&$0$ & $0$ & $0$ & $0$ & $\eps$\\}
$\ k_6\ra 0\ $&$\varepsilon$ & $-\varepsilon$ & $\varepsilon$ & $-\varepsilon$ & $\varepsilon$\\
\hline
\end{tabular}
\end{center}

\subsection{Collinear Limits}\noindent
The collinear limit is defined as two adjoining momenta $k_i$ and $k_{i+1}$, with $i+1$ mod $N$, becoming parallel. Due to cyclic symmetry, these can be chosen as $k_{N-1}$ and $k_N$,
with $k_{N-1}$ carrying the fraction $x$ of the combined momentum $k_{N-1}+k_N\to k_{N-1}$. Formally,
\be k_{N-1}\to x\, k_{N-1}\qquad,\qquad k_{N}\to (1-x)k_{N-1}\ ,\nonumber\ee
where the  momenta appearing in the limits describe the scattering of $N-1$ gluons.
{}For $N=6$, the collinear limits of scalar invariants,
written in terms of the invariants describing $N=5$ scattering, are:
\begin{center}$\bm{N=6}$\\
\begin{tabular}{|c|c|c|c|c|c|c|c|c|c|}
\hline
 & $ s_1 $ & $ s_2 $ & $ s_3 $ & $ s_4 $ & $ s_5 $ & $ s_6 $ & $ t_1 $ & $ t_2 $ & $ t_3$\\
\hline
\comment{$0$ & $(1-x)s_1$ & $s_2$ & $s_3$ & $s_4$ & $xs_5$ & $s_1$ & $xs_2+(1-x)\ s_4$ & $s_5$\\
$x\ s_1$ & $0$ & $(1-x)s_2$ & $s_3$ & $s_4$ & $s_5$ & $s_1$ & $s_2$ & $xs_3+(1-x)s_5$\\
$s_1$ & $xs_2$ & $0$ & $(1-x)s_3$ & $s_4$ & $s_5$ & $xs_4+(1-x)s_1$ & $s_2$ & $s_3$ \\
$ s_1$ & $ s_2$ & $x\ s_3$ & $0$ & $(1-x)s_4$ & $s_5$ & $s_4$ & $(1-x)s_2+xs_5$ & $s_3$\\ }
$k_{5}\to x\, k_{5}$, $\,k_{6}\to (1-x)k_{5}$ &
$ s_1$ & $ s_2$ & $s_3$ & $xs_4$ & $0$ & $(1-x)s_5$ & $s_4$ & $s_5$ & $xs_1+(1-x)s_3$\\
\hline
\end{tabular}\end{center}\vskip 3mm
The collinear limits of pseudoscalar invariants  are written below:
\begin{center}$\bm{N=6}$\\
\begin{tabular}{|c|c|c|c|c|c|c|c|c|c|}
\hline
\ &$ \epsilon_1 $ & $ \epsilon_2 $ & $ \epsilon_3 $ & $ \epsilon_4 $ & $ \epsilon_5 $ \\
\hline
$k_{5}\to x\, k_{5}$, $\,k_{6}\to (1-x)k_{5}$&$\varepsilon$ & $-x\, \varepsilon$ & $x\, \varepsilon$ & $-x\, \varepsilon$ & $x\, \varepsilon$\\
\hline
\end{tabular}
\end{center}
\section{\label{appendixB} $\bm{\ap}$ Expansions of Triple Hypergeometric\\ \hspace*{3.7cm} Functions}
Most of the $\ap$ expansions of triple hypergeometric functions \req{triple} presented in Ref.\cite{Oprisa:2005wu} apply to non-singular functions without poles, like $F_3$, see Eq.\req{expf3}.
In that case, the expansions of the integrals \req{triple} 
can be directly mapped to convergent Euler-Zagier sums. 
In this appendix we derive the expansions \req{exR1} and \req{exR2} for the singular functions $F_1$ and $F_2$, respectively.

\subsection{$\bm F_1$: Triple Hypergeometric Function with a Triple Pole}

Let us divide the integral defining $F_1$ into two parts, $I_1$ and $I_2$:
\begin{eqnarray}
\ts\FF{3,2,1}{0,0,0,0,0,0}&=&
\int_0^1\!\!\! dx\!\!\int_0^1\!\!\!dy\!\!\int_0^1\!\!\!dz\
x^{s_2-1}\ y^{t_2-1}\ z^{s_6-1}\ (1-x)^{s_3}\ (1-y)^{s_4}\ (1-z)^{s_5}\nonumber \\
&\times& (1-xy)^{t_3-s_3-s_4}\ (1-yz)^{t_1-s_4-s_5}\
(1-xyz)^{s_1+s_4-t_1-t_3}=I_1+I_2\ ,\hskip1cm 
\end{eqnarray}
with:
\begin{eqnarray}
I_1&=&\ds{\underbrace{\lf(\int_0^1\!\!\! dx\ x^{s_2-1}\ (1-x)^{s_3}\ri)}_{=\fc{\Gamma(s_2)\
  \Gamma(1+s_3)}{\Gamma(1+s_2+s_3)}}\ \ 
\underbrace{\lf(\int_0^1\!\!\!dy\!\!\int_0^1\!\!\!dz\
y^{t_2-1}\ z^{s_6-1}\ (1-y)^{s_4}\
(1-z)^{s_5}\ (1-yz)^{t_1-s_4-s_5}\ri)}_{=\fc{\Gamma(s_6)\ \Gamma(t_2)\ \Gamma(1+s_4)\ 
\Gamma(1+s_5)}{\Gamma(1+s_4+t_2)\ \Gamma(1+s_5+s_6)}\ 
\F{3}{2}\lf[{s_6,\ t_2,\ s_4+s_5-t_1\atop 1+s_5+s_6,\ 1+s_4+t_2};1\ri]}\ ,}\nonumber\\[5mm]
I_2&=&\int_0^1\!\!\! dx\!\!\int_0^1\!\!\!dy\!\!\int_0^1\!\!\!dz\
x^{s_2-1}\ y^{t_2-1}\ z^{s_6-1}\ (1-x)^{s_3}\ (1-y)^{s_4}\ (1-z)^{s_5}\ 
(1-yz)^{t_1-s_4-s_5}\nonumber \\
&&\hskip5cm\times\lf[\ (1-xy)^{t_3-s_3-s_4}\
(1-xyz)^{s_1+s_4-t_1-t_3}-1\ \ri]\ .
\end{eqnarray}
The first integral $I_1$ involves the Beta function \req{boilo} and the
hypergeometric function $\F{3}{2}$ \req{boil}. In fact, the latter integral 
is $f_1$ of Eq.\req{onefinds} with appropriate arguments 
and its $\ap$ expansion can be found in Eq.\req{rexp}. 
On the other hand, expanding the Beta function is straightforward,
so altogether we obtain:
\begin{eqnarray}
\fc{\Gamma(s_2)\ \Gamma(1+s_3)}{\Gamma(1+s_2+s_3)}=\fc{1}{s_2}-\zeta(2)\
s_3+\zeta(3)\ s_3\ (s_2+s_3) +\ldots&&,  \\[5mm]
\fc{\Gamma(s_6)\ \Gamma(t_2)\ \Gamma(1+s_4)\ 
\Gamma(1+s_5)}{\Gamma(1+s_4+t_2)\ \Gamma(1+s_5+s_6)}\ 
{\ts\F{3}{2}\lf[{s_6,t_2,s_4+s_5-t_1\atop 1+s_5+s_6,1+s_4+t_2};1\ri]}&=&\fc{1}{s_6t_2}-
\zeta(2)\ \lf(\fc{s_4}{s_6}+\fc{s_5}{t_2}  \ri)\nonumber\\
+\zeta(3)\ \lf(s_4+s_5-t_1+\fc{s_4\ (s_4+t_2)}{s_6}+\fc{s_5\ (s_5+s_6)}{t_2}\ri)&+&
\ldots\ \nonumber.
\end{eqnarray}
Hence, we obtain the following $\ap$--expansion for the integral $I_1$:
\begin{eqnarray}
I_1&=&\fc{1}{s_2s_6t_2}-\zeta(2)\ \lf(\ \fc{s_4}{s_2s_6}+\fc{s_5}{s_2t_2}+
\fc{s_3}{s_6t_2}\ \ri)\nonumber \\
&+&\zeta(3)\ \lf(\fc{s_4+s_5-t_1}{s_2}+\fc{s_4^2+s_4t_2}{s_2s_6}+
\fc{s_5^2+s_5s_6}{s_2t_2}+\fc{s_2s_3+s_3^2}{s_6t_2}\ri)+\ldots\ .
\label{cern1}
\end{eqnarray}
The second integral $I_2$ has a single pole in $s_6$ originating from $z\ra 0$ in 
the integrand.
Its expansion in $\ap$ amounts to expanding it in powers of $s_6$:
\begin{eqnarray}
s_6^{-1}&:&-(s_3+s_4-t_3)\ \int_0^1\!\!\! dx\!\!\int_0^1\!\!\!dy\int_0^1\!\!\!dz\ 
z^{s_6-1}\ \fc{\ln(1-xy)}{xy}
=\fc{(s_3+s_4-t_3)}{s_6}\ \zeta(3)\ , \nonumber \\
&&-\fc{s_3+s_4-t_3}{s_6}\int_0^1\!\!\! dx\!\!\int_0^1\!\!\!dy\ 
\fc{\ln(1-xy)}{xy}\ \lf[s_3\ \ln(1-x)+s_2\ \ln(x)+s_4\ \ln(1-y)+t_2\ \ln(y)\ri]
\nonumber\\
&+&\fc{(s_3+s_4-t_3)^2}{2\ s_6}\int_0^1\!\!\! dx\!\!\int_0^1\!\!\!dy\ \fc{\ln(1-xy)^2}{xy}
=-\fc{\zeta(4)}{4}\ \fc{(s_3 + s_4 - t_3)}{s_6}\ 
[4\ (s_2 + s_3 + s_4 + t_2) + t_3]\ ,\nonumber\\
s_6^0&:&(s_1+s_4-t_1-t_3)\ \int_0^1\!\!\! dx\!\!\int_0^1\!\!\!dy\int_0^1\!\!\!dz\ 
\fc{\ln(1-xyz)}{xyz}=-(s_1+s_4-t_1-t_3)\ \zeta(4)\ .
\label{cern}
\end{eqnarray}
Here, we have applied the following  basic Euler integrals of the type 
\req{Polyintegral}:
\begin{eqnarray}
&&
\int_0^1\!\!\! dx\!\!\int_0^1\!\!\!dy\ \ \fc{\ln x\ \ln (1-xy)}{xy}=\zeta(4)\ \ \ ,\ \ \ 
\int_0^1\!\!\! dx\!\!\int_0^1\!\!\!dy\ \ \fc{\ln(1-x) 
\ln (1-xy)}{xy}=\fc{5}{4}\ \zeta(4)\ ,\nonumber\\
&&
\int_0^1\!\!\! dx\!\!\int_0^1\!\!\!dy\ \ \fc{\ln (1-xy)^2}{xy}=\h\ \zeta(4)\ \ \ ,\ \ \ 
\int_0^1\!\!\! dx\!\!\int_0^1\!\!\!dy\ \ \fc{\ln (1-xy)}{xy}=-\zeta(3)\ ,\nonumber\\
&&
\int_0^1\!\!\! dx\!\!\int_0^1\!\!\!dy\!\!\int_0^1\!\!\!dz\ \ \ 
\fc{\ln (1-xyz)}{xyz}=-\zeta(4)\ .
\end{eqnarray}
In this way, we obtain
\be
I_2=\fc{(s_3+s_4-t_3)}{s_6}\ \zeta(3)-\fc{\zeta(4)}{4}\ \fc{(s_3 + s_4 - t_3)}{s_6}\ 
[4\ (s_2 + s_3 + s_4 + t_2) + t_3]-(s_1+s_4-t_1-t_3)\ \zeta(4)+\ldots
\label{cern2}
\ee
{}Finally, after putting together \req{cern1} and \req{cern2}, we obtain Eq.\req{exR1}.

\subsection{$\bm{F_2}$: Triple Hypergeometric Function with a Single Pole}
Here again, we divide the integral defining  $F_2$ into two parts, $I_1$ and $I_2$:
\begin{eqnarray}
\ts\FF{4,3,1}{0,-1,0,0,0,0}&=&
\int_0^1\!\!\! dx\!\!\int_0^1\!\!\!dy\!\!\int_0^1\!\!\!dz\
x^{s_2}\ y^{t_2}\ z^{s_6-1}\ (1-x)^{s_3}\ (1-y)^{s_4}\ (1-z)^{s_5}\nonumber \\
&\times& (1-xy)^{t_3-s_3-s_4-1}\ (1-yz)^{t_1-s_4-s_5}\
(1-xyz)^{s_1+s_4-t_1-t_3}=I_1+I_2\hskip1cm 
\end{eqnarray}
with:
\begin{eqnarray}
I_1&=&\ds{\underbrace{\lf(\int_0^1\!\!\! dx\ z^{s_6-1}\ (1-z)^{s_5}\ri)}_{
=\fc{\Gamma(1+s_5)\ \Gamma(s_6)}{\Gamma(1+s_5+s_6)}}\ \ 
\underbrace{\lf(\int_0^1\!\!\!dx\!\!\int_0^1\!\!\!dy\
x^{s_2}\ y^{t_2}\ (1-x)^{s_3}\
(1-y)^{s_4}\ (1-xy)^{t_3-s_3-s_4-1}\ri)}_{=\fc{\Gamma(1+s_2)\ \Gamma(1+s_3)\ 
\Gamma(1+s_4)\ \Gamma(1+t_2)\ }{\Gamma(2+s_2+s_3)\ \Gamma(2+s_4+t_2)}\ 
\F{3}{2}\lf[{1+s_2,\ 1+t_2,\ 1+s_3+s_4-t_3\atop 2+s_2+s_3,\ 2+s_4+t_2};1\ri]}\ ,}
\nonumber\\[5mm]
I_2&=&\int_0^1\!\!\! dx\!\!\int_0^1\!\!\!dy\!\!\int_0^1\!\!\!dz\
x^{s_2}\ y^{t_2}\ z^{s_6-1}\ (1-x)^{s_3}\ (1-y)^{s_4}\ (1-z)^{s_5}\ 
(1-xy)^{t_3-s_3-s_4-1}
\nonumber \\
&&\hskip5cm\times\lf[\ (1-yz)^{t_1-s_4-s_5}\ (1-xyz)^{s_1+s_4-t_1-t_3}-1\ \ri]\ .
\end{eqnarray}
The first integral $I_1$ involves the Beta function \req{boilo} and the
hypergeometric function $\F{3}{2}$ \req{boil}. In fact, the latter integral 
is $f_2$ of Eq.\req{onefinds} with appropriate arguments
and its $\ap$ expansion can be found in Eq.\req{rexp}. 
On the other hand, expanding the Beta function is straightforward,
so altogether we obtain:
\begin{eqnarray}
\fc{\Gamma(s_6)\ \Gamma(1+s_5)}{\Gamma(1+s_5+s_6)}&=&\fc{1}{s_6}-\zeta(2)\
s_5+\zeta(3)\ s_5\ (s_5+s_6) +\ldots\ ,  \\[5mm]
\fc{\Gamma(1+s_2)\ \Gamma(1+s_3)\ 
\Gamma(1+s_4)\ \Gamma(1+t_2)\ }{\Gamma(2+s_2+s_3)\ \Gamma(2+s_4+t_2)}&&
\hskip-0.7cm{\ts\F{3}{2}\lf[{1+s_2,\ 1+t_2,\ 1+s_3+s_4-t_3\atop 2+s_2+s_3,\ 2+s_4+t_2};1\ri]}
\nonumber\\
&=&\zeta(2)-\zeta(3)\ \lf(\ s_2+s_3+s_4+t_2+t_3\ \ri)+\ldots\ .\nonumber
\end{eqnarray}
Hence, we obtain the following $\ap$--expansion for the integral $I_1$:
\be
I_1=\fc{\zeta(2)}{s_6}-\zeta(3)\ \fc{s_2+s_3+s_4+t_2+t_3}{s_6}+\ldots\ .
\label{Cern1}
\ee
The integrand of the second integral $I_2$ remains finite for $z\ra 0$.
Up to the first leading order, it involves the following finite sub-integrals:
$$\int_0^1\!\!\! dx\!\!\int_0^1\!\!\!dy\!\!\int_0^1\!\!\!dz\ \ \ 
\fc{\ln (1-yz)}{z\ (1-xy)}=-\fc{5}{4}\ \zeta(4)\ \ \ ,\ \ \ 
\int_0^1\!\!\! dx\!\!\int_0^1\!\!\!dy\!\!\int_0^1\!\!\!dz\ \ \ 
\fc{\ln (1-xyz)}{(1-xy)\ z}=-\fc{3}{4}\ \zeta(4)\ .$$
With this information we obtain
\be
I_2=\fc{5}{4}\ (s_4+s_5-t_1)\ \zeta(4)-\fc{3}{4}\ (s_1+s_4-t_1-t_3)\ \zeta(4)+\ldots\ .
\label{Cern2}
\ee
After putting together \req{Cern1} and \req{Cern2} we obtain Eq.\req{exR2}.

Finally, the four functions $F_3,F_4,F_5$ and $F_6$ do not contain any poles in the kinematic
invariants \req{in}. Hence their $\ap$ expansions can be obtained by the methods described
in \cite{Oprisa:2005wu}, \ie by evaluating  the relevant  Euler-Zagier sums.

\section{\label{appendixC}Functions $\bm{V^{(6)}}$ and
$\bm{P_i^{(6)}}$ Expressed in the Basis \\ \hspace*{3.7cm} ${\large \bm{F_k,\ k=1,...,6}}$}
The functions $P_i^{(6)}$ and $V^{(6)}$ governing the
six-gluon  MHV amplitude \req{m6} are expressed in Eqs.\req{ps6} and \req{vs6} in terms of certain generalized hypergeometric integrals, in the notation of Eq.\req{triple}.
In Section IV B, we introduced a basis of six functions, see Eq.\req{FUNCS}, which is very convenient for studying cyclic properties and low-energy limits.
The integrals appearing in Eqs.\req{ps6} and \req{vs6} can be expressed in this basis by using the relations obtained in Ref.\cite{Oprisa:2005wu} as a combined result of partial integrations, use of world-sheet supersymmetry {\it etc.} The functions that enter Eq.\req{ps6} are:   
\begin{eqnarray}
\ts\FF{4,4,3}{0,0,0,-1,-1,0}&=&-F_3+F_6\ ,\nnn\\
s_1\ \ts\FF{4,3,2}{0,-1,0,-1,-1,0}&=&s_6\ (F_2+F_4) 
+ (s_2 + s_5 - t_1 - t_2)\ (F_3+F_4-F_5) \nnn\\
&+& (s_1 - s_5 - s_6 + t_2)\ F_6\ , \nnn\\
s_2\ \ts\FF{3,3,2}{0,-1,0,0,-1,0}&=&s_6\ (F_2-F_3) +
(s_2 - s_3 + s_5 -t_1 + t_3)\  (F_3+F_4-F_6)\nnn\\
&-&(s_1 - s_3 + s_5 - t_1)\ F_5+ (s_1 + s_2 - t_1)\ F_6\ ,\nnn\\
s_3\ \ts\FF{4,3,2}{-1,0,0,0,-1,0}&=&s_6\ (F_2-F_3) + (s_1+s_2-t_1)\ F_3-s_3\ (F_3-F_6) \nnn\\ 
&-& (s_1 - s_3 + s_5 - t_1)\ F_5+ (s_4 + s_5 -t_1)\ (F_3+F_4-F_6)\ \ ,\nnn\\
s_4\ \ts\FF{4,3,2}{0,0,-1,-1,0,0}&=&s_6\ F_2 + (s_4 -s_5- s_6 +t_2)\ F_3+
(s_4 + s_5 - t_1)\ F_4\nnn\\
&+& (s_1 - s_3 + s_5 - t_1)\ (F_3-F_5)\ ,\nnn\\
s_5\ \ts\FF{4,3,2}{0,-1,0,0,0,-1}&=&s_6\ F_2+(s_1 + s_4 - t_1 - t_3)\  (F_3-F_5)+
(s_4 + s_5 - t_1)\ F_4\ .\nnn
\end{eqnarray}
The additional function that enters Eq.\req{vs6} is: 
\begin{eqnarray}
s_2s_5\ \ts\FF{3,2,2}{0,0,0,0,0,-1}&=&s_2s_6\ F_1 -
s_2\ (s_1-s_5-t_3)\ F_3\nnn \\
&+&(s_4+ s_5- t_1)\ [\ s_6\ (F_2-F_3)-(s_3-s_5+t_1-t_3)\ (F_3+F_4)\nnn \\
&-&(s_1-s_3+s_5-t_1)\ F_5+(s_1+s_3-s_5-t_3)\ F_6\ ]\ .\nnn
\end{eqnarray}

\bibliography{gluons}% Produces the bibliography via BibTeX.

\begin{thebibliography}{36}
\bibitem[{cyc()}]{cyc}
\bibinfo{note}{In discussing cyclic transformations, we always take into
  account the $\mathbb{Z}_N$-invariant momentum conservation constraint
  $\sum_{i=1}^{i=N}k_i=0$}.
\bibitem[{app({\natexlab{b}})}]{appa2}
\bibinfo{note}{The soft and collinear limits of all invariants are given in
  Appendix A}.
\end{thebibliography}
\end{document}